\documentclass[aps,pra,amssymb,superscriptaddress,10pt,tightenlines]{revtex4-2}

\usepackage[english]{babel}
\usepackage{graphicx}
\usepackage{dcolumn}
\usepackage{bm}
\usepackage{color}
\usepackage{subfigure}
\usepackage[ansinew]{inputenc}
\usepackage{amsfonts}
\usepackage{amsmath}
\usepackage{amssymb}
\usepackage{subfigure}
\usepackage{bbm}
\usepackage{tikz}

\usepackage[colorlinks=true,linkcolor=red,urlcolor=red,citecolor=magenta,pdfusetitle]{hyperref}

\usepackage[braket]{qcircuit}
\usepackage{braket}

\usepackage{mathrsfs}
\usepackage{dsfont}

\begin{document}



\begin{center}
\title*{\Large{Transfer entropy and O-information to detect grokking in tensor network multi-class classification problems}}
\end{center}

\author{Domenico Pomarico$^\dagger$}%
\affiliation{%
	Dipartimento di Fisica, Universit\`a di Bari, I-70126 Bari, Italy
}%
\affiliation{
Istituto Nazionale di Fisica Nucleare, Sezione di Bari, I-70126 Bari, Italy}
\altaffiliation{These authors contributed equally to this work}

\author{Roberto Cilli$^\dagger$}%
\affiliation{%
	Dipartimento di Fisica, Universit\`a di Bari, I-70126 Bari, Italy
}%
\altaffiliation{These authors contributed equally to this work}

\author{Alfonso Monaco}%
\email{alfonso.monaco@ba.infn.it}
\affiliation{%
	Dipartimento di Fisica, Universit\`a di Bari, I-70126 Bari, Italy
}%
\affiliation{
Istituto Nazionale di Fisica Nucleare, Sezione di Bari, I-70126 Bari, Italy}

\author{Loredana Bellantuono}%
\affiliation{
Istituto Nazionale di Fisica Nucleare, Sezione di Bari, I-70126 Bari, Italy}
\affiliation{%
	Dipartimento di Biomedicina Traslazionale e Neuroscienze (DiBraiN), Universit\`a degli Studi di Bari, Bari, I-70124, Italy
}%

\author{Marianna La Rocca}%
\affiliation{%
	Dipartimento di Fisica, Universit\`a di Bari, I-70126 Bari, Italy
}%
\affiliation{
Istituto Nazionale di Fisica Nucleare, Sezione di Bari, I-70126 Bari, Italy}

\author{Tommaso Maggipinto}%
\affiliation{%
	Dipartimento di Fisica, Universit\`a di Bari, I-70126 Bari, Italy
}%
\affiliation{
	Istituto Nazionale di Fisica Nucleare, Sezione di Bari, I-70126 Bari, Italy}

\author{Giuseppe Magnifico}%
\affiliation{%
	Dipartimento di Fisica, Universit\`a di Bari, I-70126 Bari, Italy
}%
\affiliation{
	Istituto Nazionale di Fisica Nucleare, Sezione di Bari, I-70126 Bari, Italy}

\author{Marlis Ontivero Ortega}%
\affiliation{%
	Dipartimento di Fisica, Universit\`a di Bari, I-70126 Bari, Italy
}%

\author{Ester Pantaleo}%
\affiliation{%
	Dipartimento di Fisica, Universit\`a di Bari, I-70126 Bari, Italy
}%
\affiliation{
	Istituto Nazionale di Fisica Nucleare, Sezione di Bari, I-70126 Bari, Italy}
	
\author{Sabina Tangaro}%
\affiliation{
	Istituto Nazionale di Fisica Nucleare, Sezione di Bari, I-70126 Bari, Italy}
\affiliation{%
	Dipartimento Di Scienze Del Suolo, Della Pianta e Degli Alimenti, Universit\`a degli Studi di Bari, Bari, I-70125, Italy
}%

\author{Sebastiano Stramaglia}%
\affiliation{%
	Dipartimento di Fisica, Universit\`a di Bari, I-70126 Bari, Italy
}%
\affiliation{
	Istituto Nazionale di Fisica Nucleare, Sezione di Bari, I-70126 Bari, Italy}
	
\author{Roberto Bellotti}%
\affiliation{%
	Dipartimento di Fisica, Universit\`a di Bari, I-70126 Bari, Italy
}%
\affiliation{
	Istituto Nazionale di Fisica Nucleare, Sezione di Bari, I-70126 Bari, Italy}
	
\author{Nicola Amoroso}%
\affiliation{
	Istituto Nazionale di Fisica Nucleare, Sezione di Bari, I-70126 Bari, Italy}
\affiliation{%
	Dipartimento di Farmacia-Scienze del Farmaco, Universit\`a degli Studi di Bari, Bari, I-70125, Italy
}%

\begin{abstract}
	Quantum-enhanced machine learning, encompassing both quantum algorithms and quantum-inspired classical methods such as tensor networks, offers promising tools for extracting structure from complex, high-dimensional data. In this work, we study the training dynamics of Matrix Product State (MPS) classifiers applied to three-class problems, using both fashion MNIST and hyperspectral satellite imagery as representative datasets. We investigate the phenomenon of grokking, where generalization emerges suddenly after memorization, by tracking entanglement entropy, local magnetization, and model performance across training sweeps. Additionally, we employ information-theory tools to gain deeper insights: transfer entropy is used to reveal causal dependencies between label-specific quantum masks, while O-information captures the shift from synergistic to redundant correlations among class outputs. Our results show that grokking in the fashion MNIST task coincides with a sharp entanglement transition and a peak in redundant information, whereas the overfitted hyperspectral model retains synergistic, disordered behavior. These findings highlight the relevance of high-order information dynamics in quantum-inspired learning and emphasize the distinct learning behaviors that emerge in multiclass classification, offering a principled framework to interpret generalization in quantum machine learning architectures.
\end{abstract}

\maketitle

\pagecolor{white}

\def\thefootnote{\ddagger}\footnotetext{These authors contributed equally to this work}\def\thefootnote{\arabic{footnote}}

\section{Introduction}

The convergence of quantum physics, machine learning, and computer science is giving rise to a new computational paradigm with the potential to reshape how information is processed and learned. Quantum machine learning (QML) stands at the forefront of this development, with the aim of exploiting the unique properties of quantum mechanics to improve learning algorithms by encoding data into quantum states and elaborating them through quantum circuits \cite{arrazola, banchiscience, chinaphotonics, vakili2024, benedetti, hibat, Gili_2023, qgeneralize, holmes2024, Peters2023generalization, bowles2023contextuality, eisert2024, Pomarico2025, breastQML}. However, current quantum hardware is still facing issues such as noise, decoherence, and limited circuit depth, which impose serious constraints on the scalability and reliability of purely quantum models \cite{PRXQuantum, ibm_dqpt, ibm_mit}.

To address these limitations, hybrid quantum-classical approaches have gained prominence. These strategies incorporate quantum components within a predominantly classical framework, or emulate quantum-like behaviors using classical algorithms. Among the most successful tools in this direction are tensor network methods, which originate from condensed matter physics and quantum many-body theory. Specifically, Matrix Product States (MPS) have proven highly effective for data-driven tasks: they provide an efficient way to represent high-dimensional data with limited entanglement and allow for tracking learning mechanism via local observables, making them well suited for scalable quantum-inspired machine learning \cite{MPScircuit, Rudolph_2024, Miller_2024, schuhmacher2024hybridtreetensornetworks, Khosrojerdi_2025, stoudenmire2017supervisedlearningquantuminspiredtensor, Huggins_2019, Felser_2021, Dborin_2022, Ballarin2023entanglemententropy, Collura2021, PhysRevX8, glasser2020, Gallego2022, PhysRevB, chen2023machinelearningtreetensor, pomarico2025grokkingentanglementtransitiontensor}.

One central challenge in remote sensing tasks, such as land cover classification by means of hyper-spectral images, lies in managing the high-dimensional feature space while preserving relevant spectral correlations. In such settings, tensor network models like MPS allow us to control the expressivity and entanglement of the model, navigating the trade-off between overfitting and generalization \cite{stoudenmire2017supervisedlearningquantuminspiredtensor, Huggins_2019, Felser_2021, Dborin_2022, Ballarin2023entanglemententropy, Collura2021, PhysRevX8, glasser2020, Gallego2022, PhysRevB, chen2023machinelearningtreetensor, pomarico2025grokkingentanglementtransitiontensor}. Moreover, the phenomenon of grokking, a delayed yet sudden improvement in generalization performance after the model fits the training data, is particularly relevant in over-parameterized regimes \cite{pomarico2025grokkingentanglementtransitiontensor,power2022grokkinggeneralizationoverfittingsmall, liu2022understandinggrokkingeffectivetheory, liu2023grokkingcompressionnonlinearcomplexity, miller2024grokkingneuralnetworksempirical, varma2023explaininggrokkingcircuitefficiency, huang2024unifiedviewgrokkingdouble}: a competition arises between fast, memorizing modes and slower, generalizing dynamics \cite{varma2023explaininggrokkingcircuitefficiency, huang2024unifiedviewgrokkingdouble,montanari,seroussi}, which can lead to transitions in the information structure of the learned representation in both quantum and classical models \cite{pomarico2025grokkingentanglementtransitiontensor,rubin2024grokkingorderphasetransition, clauw2024informationtheoreticprogressmeasuresreveal}.

To quantitatively analyze these transitions, we adopt an information-theoretic perspective \cite{varley} rooted in two complementary tools. First, we apply transfer entropy \cite{schreiber_te,faes}, a directional and time-resolved measure of information flow, to track causal relationships between different components (e.g., label-specific quantum masks) of the trained MPS during learning. This approach reveals asymmetric causal dependencies and the emergence of hierarchical information channels during training. Second, we leverage the O-information \cite{rosas}, a high-order generalization of mutual information, to quantify whether the interaction among multiple outputs is dominated by redundancy (shared, overlapping information) or synergy (complementary, distributed information). Positive O-information indicates redundant correlations, while negative values highlight synergistic structure.

These tools provide a refined lens through which to observe learning phase transitions in quantum machine learning models. In particular, we show how a transition in entanglement entropy correlates with changes in information structure, as encoded by O-information and transfer entropy, across different tasks. While this framework has broad applicability, we focus on two representative datasets: the widely used fashion MNIST dataset and a challenging hyperspectral satellite image dataset for land cover classification. The former exhibits clear signs of grokking, including a transition from synergistic to redundant inter-label information and the emergence of causal mask interactions. The latter remains trapped in a regime dominated by synergy and weak causal structure, consistent with overfitting and poor generalization. Among the three fashion MNIST classes, i.e. dress, sneaker and bag, the last acts as a confounding class, introducing controlled difficulty in feature separation. In contrast, the PRISMA dataset, with labels such as grapevine, olive tree, and cropland, presents intrinsic noise and label uncertainty, often resulting in limited generalization. This dual setting enables a comparative study of grokking under both learnable and noisy conditions, and supports the analysis of emergent causal and high-order information structures.

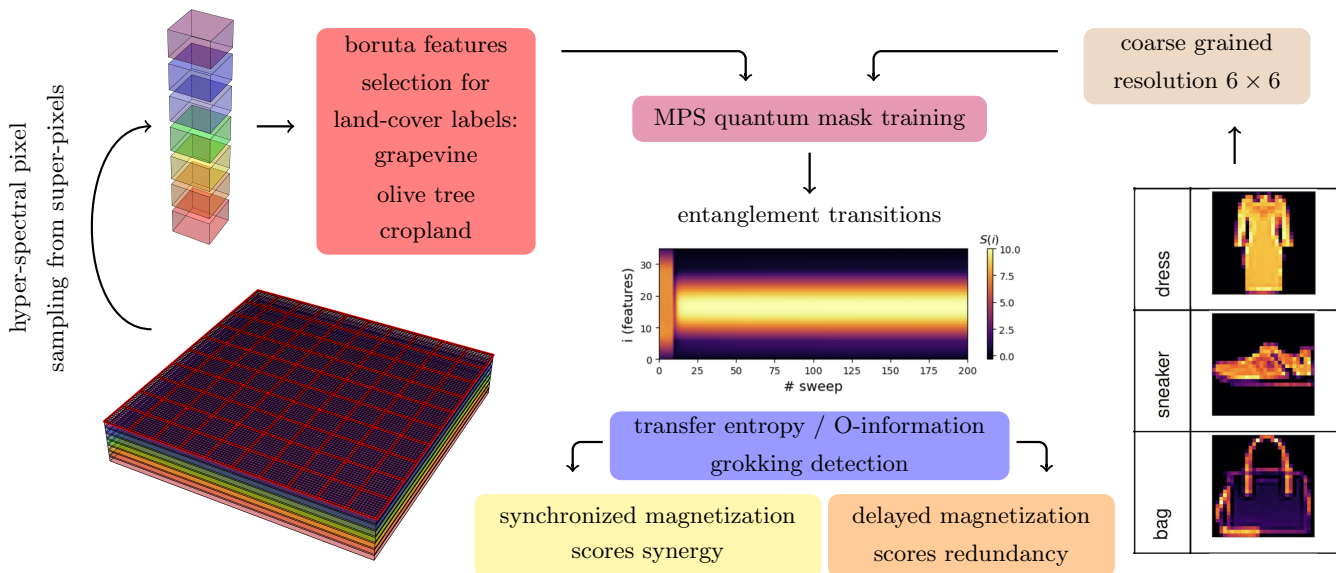
\begin{figure}[t!]
\centering
\begin{tikzpicture}
\node[inner sep=0pt] (russell) at (-0.45,-5) {\includegraphics[clip, width = 0.3\linewidth{}]{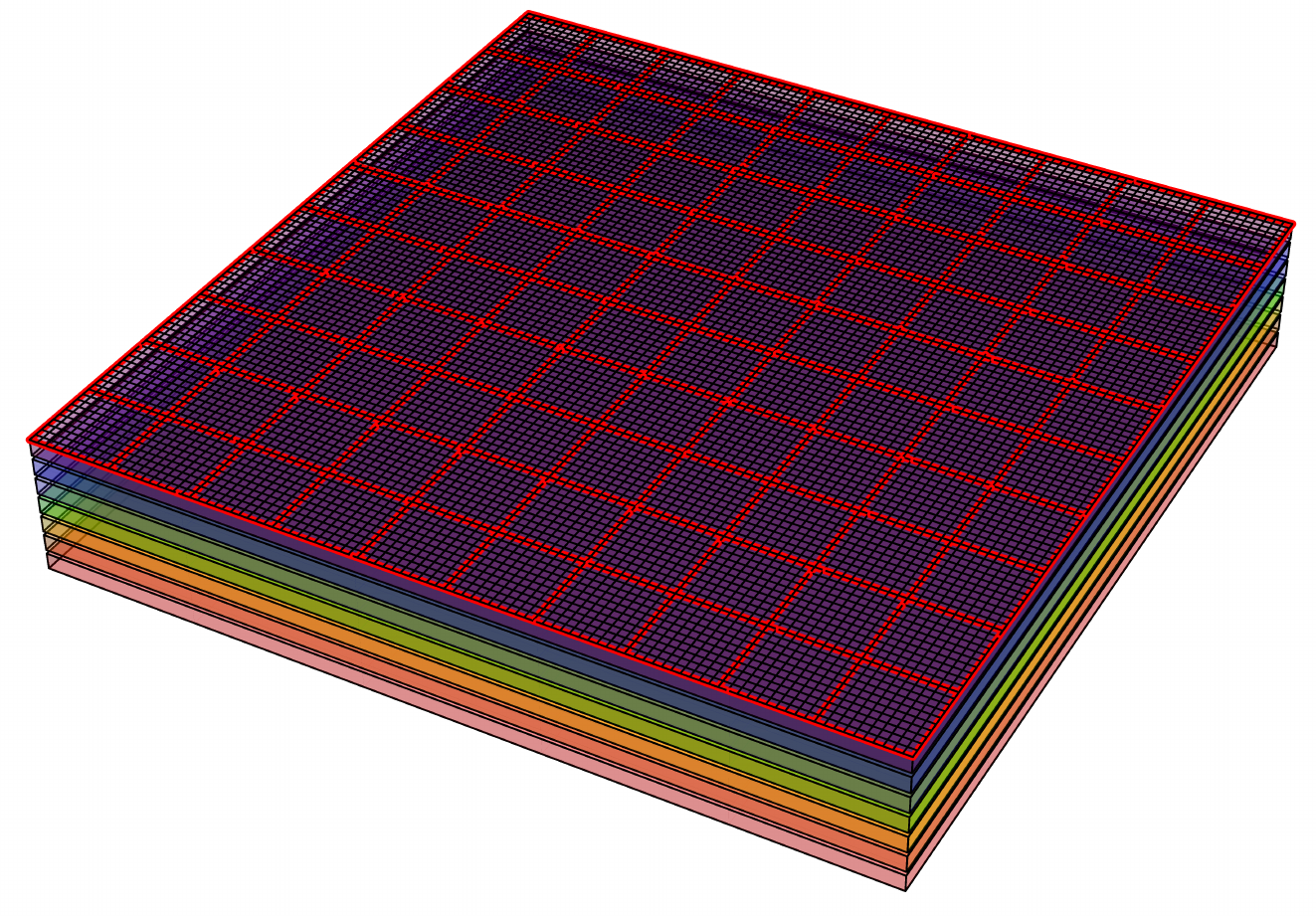}};
\draw[->,thick] (-2.4,-3.7) to [bend right=270] (-2.4,-1);
\node[] (t8) at (-4.15,-2.35) {\rotatebox{90}{hyper-spectral pixel}};
\node[] (t8) at (-3.65,-2.35) {\rotatebox{90}{sampling from super-pixels}};
\draw[->, line width=0.3mm] (-1,-1) -- (-0.5,-1);
\node[inner sep=0pt] (russell2) at (-1.75,-1) {\includegraphics[clip, width = 0.05\linewidth{}]{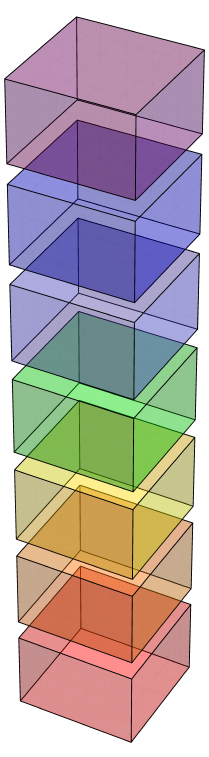}};
\draw[rounded corners,red!50,fill] (-0.2, 0.3) rectangle (2.7, -2.7);
\node[] (t8) at (1.25,0.1) {boruta features};
\node[] (t8) at (1.25,-0.4) {selection for};
\node[] (t8) at (1.25,-0.9) {land-cover labels:};
\node[] (t8) at (1.25,-1.4) {grapevine};
\node[] (t8) at (1.25,-1.9) {olive tree};
\node[] (t8) at (1.25,-2.4) {cropland};
\node[inner sep=0pt] (russell) at (12,-4.25) {\includegraphics[clip, width = 0.15\linewidth{}]{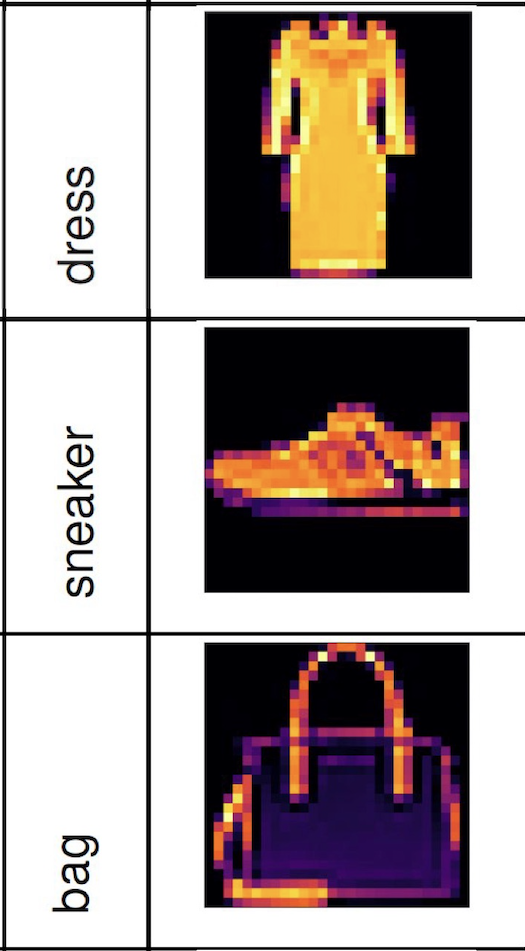}};
\draw[rounded corners,brown!30,fill] (10, 0.3) rectangle (13, -0.7);
\draw[<-, line width=0.3mm] (12,-1) -- (12,-1.5);
\node[] (t8) at (11.5,0.1) {coarse grained};
\node[] (t8) at (11.5,-0.4) {resolution $6\times 6$};
\draw[rounded corners=3pt,line width=0.3mm,->]
  (3.05,0) -- (5.5,0) |- (5.5,-0.4);
\draw[rounded corners=3pt,line width=0.3mm,->]
  (9.65,0) -- (7.2,0) |- (7.2,-0.4);
\draw[rounded corners,purple!40,fill] (4, -0.6) rectangle (8.7, -1.2);
\node[] (t8) at (6.35,-0.9) {MPS quantum mask training};
\draw[->, line width=0.3mm] (6.35,-1.35) -- (6.35,-1.9);
\node[] (t8) at (6.35,-2.2) {entanglement transitions};
\node[inner sep=0pt] (russell) at (6.5,-3.5) {\includegraphics[clip, width = 0.3\linewidth{}]{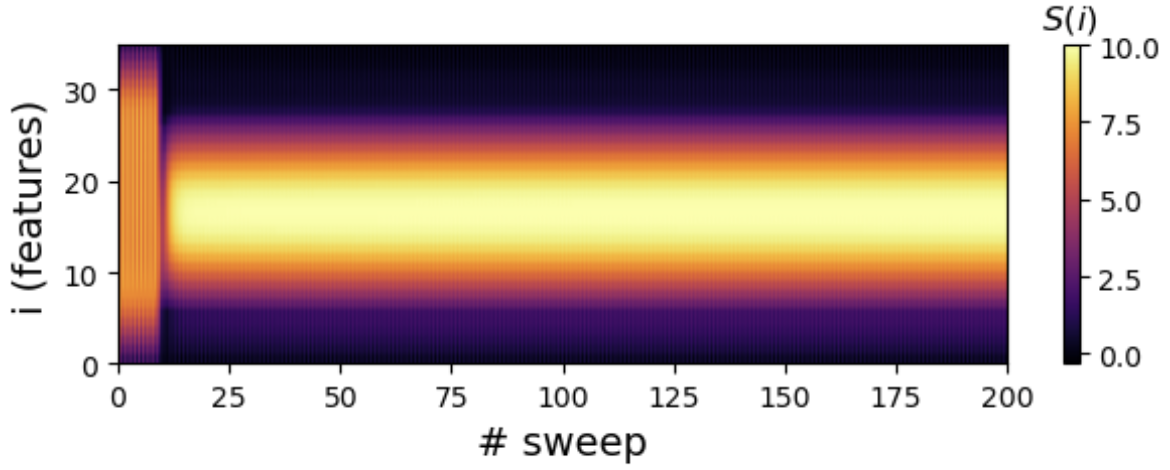}};
\draw[rounded corners,yellow!40,fill] (1.9, -5.9) rectangle (6.5, -7);
\draw[rounded corners,orange!40,fill] (6.6, -5.9) rectangle (10.4, -7);
\node[] (t8) at (8.5,-6.2) {delayed magnetization};
\node[] (t9) at (8.5,-6.7) {scores redundancy};
\node[] (t8) at (4.2,-6.2) {synchronized magnetization};
\node[] (t9) at (4.2,-6.7) {scores synergy};
\draw[rounded corners,blue!40,fill] (3.7, -4.8) rectangle (9, -5.7);
\node[] (t8) at (6.35,-5) {transfer entropy / O-information};
\node[] (t8) at (6.35,-5.5) {grokking detection};
\draw[rounded corners=3pt,line width=0.3mm,->]
  (3.6,-5.2) -- (3.2,-5.2) |- (3.2,-5.6);
\draw[rounded corners=3pt,line width=0.3mm,->]
  (9.1,-5.2) -- (9.5,-5.2) |- (9.5,-5.6);
\end{tikzpicture}
    \caption{Workflow representation for the considered multi-class classification problems pair, hyper-spectral images for land-cover and the benchmark fashion MNIST dataset. \label{fig:flowchart}}
\end{figure}

Our decision to focus on three-class classification problems is motivated by both theoretical and practical considerations related to the study of emergent correlations and learning dynamics in quantum-inspired models. In particular, the use of O-information requires at least three interacting variables to be meaningfully defined. Applying O-information to the output classification scores allows us to probe how information is distributed across the three label subspaces, capturing collective behaviors that are invisible in binary settings. Moreover, the presence of three classes enables the emergence of non-trivial causal relationships among the corresponding MPS masks, whose magnetizations evolve during training. By computing pairwise transfer entropies between these masks, we can track how the activation of one class may temporally influence others, uncovering latent structure in the learning process. This is particularly important for understanding grokking transitions, where generalization is not only a matter of performance but also of reorganization in the causal and informational architecture of the model. Therefore, the three-class setup provides a minimal yet sufficient setting to jointly analyze higher-order correlations and directional interactions, offering deeper insights into the internal dynamics of tensor network learning.

In summary, this study provides a detailed account of the learning dynamics in quantum-inspired MPS models for fashion MNIST and satellite image classification, revealing how causal and high-order information-theoretic indicators serve as robust probes of generalization, structure, and phase transitions in quantum machine learning. In this work, we focus on three-class classification problems presented in Section \ref{sec2}, as a natural extension beyond the well-studied binary case. Although two-class quantum classification has been thoroughly analyzed, particularly in the context of tensor network models, as discussed in \cite{pomarico2025grokkingentanglementtransitiontensor}, the multiclass setting introduces richer structure, more complex decision boundaries, and novel information-theory behaviors that are described in Section \ref{sec3}. A comprehensive discussion of the observed learning dynamics for the targeted classification tasks is presented in Section \ref{sec4}.

\section{Materials and Methods} \label{sec2}

The characterizing workflow of this study is represented in Figure \ref{fig:flowchart}, which presents a comparative approach outlining the data preprocessing and learning pipeline for two distinct three-class classification tasks: one based on hyperspectral satellite imagery for land cover classification, and the other using fashion MNIST images as a structured vision benchmark. The PRISMA hyper-spectral images are partitioned in super-pixels $15 \times 15$ in order to sample single pixels beyond a typical correlation length. These 3D image cubes consist of hundreds of contiguous spectral bands. From this spectral stack, Boruta feature selection is applied to identify the most relevant spectral features for distinguishing the land cover classes: grapevine, olive tree, cropland. This selected feature set endowed with 43 elements is then used to train MPS quantum masks. Fashion MNIST images are coarse-grained to a lower resolution of $6 \times 6$ pixels to match a manageable input size of the tensor network model. The three considered classes are: dress, sneaker, bag. These low-resolution inputs are also fed into the MPS quantum mask training pipeline. During training the system undergoes entanglement transitions, characterized by changes in the entanglement entropy along the qubit chain. These transitions are analyzed using two complementary information-theory tools: transfer entropy, quantifying directional, pairwise causal information flow between spins, and O-information, capturing high-order information structure, revealing synergy or redundancy in the collective scores dynamics. For the land cover problem, the trained system exhibits a synchronized magnetization across labels and a dominance of synergistic interactions among the three output scores. For the fashion MNIST task, the system shows delayed magnetization transitions (with one label magnetizing before the others) and a stronger presence of redundant scores information, especially corresponding to grokking.

\subsection{PRISMA hyper-spectral dataset}

For the land cover classification task, we employ hyperspectral imagery acquired by the PRISMA (PRecursore IperSpettrale della Missione Applicativa) satellite, an Earth observation mission developed by the Italian Space Agency (ASI). The area of interest is primarily located in the province of Barletta-Andria-Trani, in the Apulia region of southern Italy. This region features a heterogeneous agricultural landscape, including grapevine, olive trees, and cropland, making it a suitable testbed for remote sensing-based land cover analysis.

PRISMA provides high-resolution hyperspectral data across two spectral ranges: the Visible and Near-Infrared (VNIR) domain, covering wavelengths from 400 to $1010 \ nm$ with a spectral resolution of approximately $2.77 \ nm$, and the Short-Wave Infrared (SWIR) domain, spanning 920 to $2505 \ nm$ with a resolution of around $10 \ nm$, yielding over 240 contiguous channels.

Our analysis is based on Level 0 PRISMA data, which are raw satellite acquisitions subsequently processed to align with high-quality land cover annotations. A key step in our workflow involves resolution enhancement, where the native $30 \ m \times 30 \ m$ spatial resolution of the hyperspectral data is upsampled to $10 \ m \times 10 \ m$.

The upsampled images are then co-registered with land cover labels provided by AGEA (Agenzia per le Erogazioni in Agricoltura), ensuring accurate spatial correspondence between spectral data and class annotations. This integration supports supervised learning and evaluation of classification models under real-world conditions typical of satellite-based agricultural monitoring.

\subsection{Fashion MNIST dataset}

The fashion MNIST dataset serves as a well-established benchmark in the machine learning community, often used to assess the performance of classification algorithms under realistic yet manageable conditions. It comprises $60,000$ grayscale images of clothing items, each with a resolution of $28 \times 28$ pixels, uniformly distributed across 10 distinct classes: T-shirt/top, trouser, pullover, dress, coat, sandal, shirt, sneaker, bag, and ankle boot. In order to construct a dataset with a controlled cardinality and to mitigate potential confounding effects on grokking dynamics, particularly those arising from dataset size, we extract a stratified random sample corresponding to $10\%$ of the original data. This subset is subsequently split evenly into training and test sets, ensuring both class balance and statistical robustness.

To further simplify the problem while retaining its essential structure, we focus on a reduced three-class classification task involving the categories dress, sneaker, and bag. This specific selection is motivated both by visual distinguishability at low resolution and by computational constraints, as detailed in Figure \ref{fig:flowchart}. To make the simulations tractable, all images are downsampled to a compact $6 \times 6$ pixel format using an appropriate interpolation method that preserves global shape information. Despite this aggressive compression, key visual features remain identifiable, particularly for the sneaker class, which is characterized by high-contrast pixel patterns in the first and last rows, enabling relatively robust recognition even in the low-dimensional regime.

For subsequent modeling, each image is linearized into a one-dimensional sequence by flattening the pixel matrix row-wise, from top to bottom. This transformation allows us to interpret the data as configurations on a one-dimensional lattice, a setup providing a controlled yet nontrivial environment for studying emergent learning phenomena as grokking.

\subsection{Tensor network}

Tensor network machine learning capitalizes on the ability of tensor networks to compactly represent high-dimensional data. These techniques originate from quantum many-body physics and quantum information theory, where they have proven essential for analyzing both equilibrium and dynamical properties of complex quantum systems \cite{PhysRevLett.69.2863, RevModPhys.77.259, SCHOLLWOCK201196}. Among the most widely adopted architectures for machine learning are Matrix Product States (MPS), known in applied mathematics as tensor trains \cite{Tucker1966, hackbusch2019, delathauwer2000, oseledets2011}. MPS provide an efficient decomposition of large, multi-dimensional tensors into a sequential chain of low-rank tensors, dramatically reducing the number of parameters while preserving expressive power.

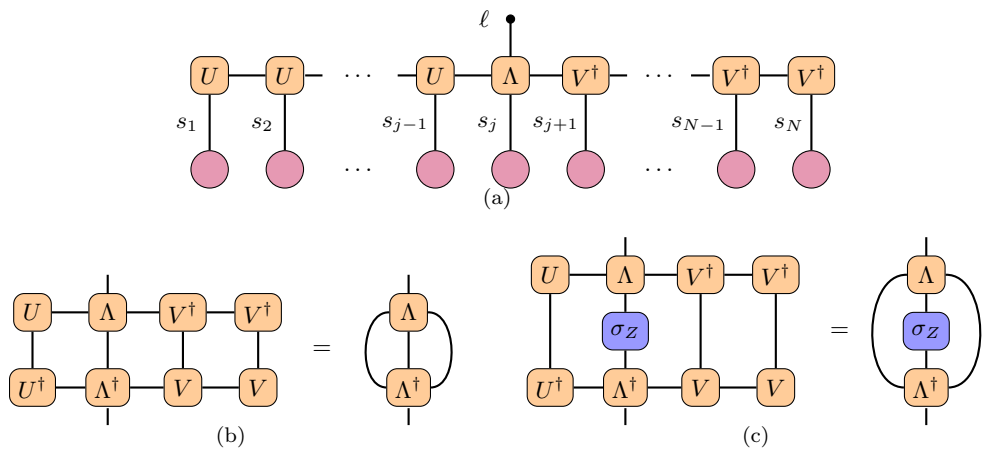
\begin{figure}[b!]
\begin{center}
\begin{tabular}{c}
    \subfigure[]{\centering
    \begin{tikzpicture}
    \node[draw, rounded corners, fill=orange!40, shape=rectangle, minimum width=0.5cm, minimum height = 0.5cm] (v0z) at (-1,0.5) {$U$};
    \node[draw, rounded corners, fill=orange!40, shape=rectangle, minimum width=0.5cm, minimum height = 0.5cm] (v0b) at (0,0.5) {$U$};
    \node[draw, rounded corners, fill=orange!40, shape=rectangle, minimum width=0.5cm, minimum height = 0.5cm] (v0a) at (2,0.5) {$U$};
    \node[draw, rounded corners, fill=orange!40, shape=rectangle, minimum width=0.5cm, minimum height = 0.5cm] (v0c) at (3,0.5) {$\Lambda$};
    \node[draw, rounded corners, fill=orange!40, shape=rectangle, minimum width=0.5cm, minimum height = 0.5cm] (v2a) at (4,0.5) {$V^\dagger$};
    \node[draw, rounded corners, fill=orange!40, shape=rectangle, minimum width=0.5cm, minimum height = 0.5cm] (v2b) at (6,0.5) {$V^\dagger$};
    \node[draw, rounded corners, fill=orange!40, shape=rectangle, minimum width=0.5cm, minimum height = 0.5cm] (v2z) at (7,0.5) {$V^\dagger$};
    \node[draw, fill=purple!40, shape=circle, inner sep=5 pt] (v00z) at (-1,-0.75) {};
    \node[draw, fill=purple!40, shape=circle, inner sep=5 pt] (v00) at (0,-0.75) {};
    \node[draw, fill=purple!40, shape=circle, inner sep=5 pt] (v00a) at (2,-0.75) {};
    \node[draw, fill=purple!40, shape=circle, inner sep=5 pt] (v00b) at (3,-0.75) {};
    \node[draw, fill=purple!40, shape=circle, inner sep=5 pt] (v22a) at (4,-0.75) {};
    \node[draw, fill=purple!40, shape=circle, inner sep=5 pt] (v22c) at (6,-0.75) {};
    \node[draw, fill=purple!40, shape=circle, inner sep=5 pt] (v22z) at (7,-0.75) {};
    \draw [thick] (v0b) -- (v00)
    (v0c) -- (v00b)
    (v0c) -- (v2a)
    (v0c) -- (v0a)
    (v2a) -- (v22a)
    (v0a) -- (v00a)
    (v0z) -- (v00z)
    (v0b) -- (v0z)
    (v2b) -- (v2z)
    (v2z) -- (v22z)
    (v2b) -- (v22c);
    \draw [thick] (3,0.75) -- (3,1.25);
    \draw [thick] (0.27,0.5) -- (0.5,0.5);
    \draw [thick] (1.5,0.5) -- (1.75,0.5);
    \draw [thick] (4.32,0.5) -- (4.55,0.5);
    \draw [thick] (5.4,0.5) -- (5.65,0.5);
    \node [] (t) at (5,0.5) {$\dots$};
    \node [] (td) at (5,-0.75) {$\dots$};
    \node [] (tb) at (1,0.5) {$\dots$};
    \node [] (tbd) at (1,-0.75) {$\dots$};
    \draw[black,fill=black] (3,1.25) circle (.4ex);
    \node [] (t3) at (2.65,1.25) {$\ell$};
    \node [] (t2) at (-1.3,-0.15) {$s_1$};
    \node [] (t2) at (-0.3,-0.15) {$s_2$};
    \node [] (t2) at (1.6,-0.15) {$s_{j-1}$};
    \node [] (t2) at (2.7,-0.15) {$s_j$};
    \node [] (t2) at (3.6,-0.15) {$s_{j+1}$};
    \node [] (t6) at (5.5,-0.15) {$s_{N-1}$};
    \node [] (t6) at (6.7,-0.15) {$s_N$};
    \end{tikzpicture}} \\
    \subfigure[]{\centering
    \begin{tikzpicture}
    \node[draw, rounded corners, fill=orange!40, shape=rectangle, minimum width=0.5cm, minimum height = 0.5cm] (v0b) at (2,0.5) {$U$};
    \node[draw, rounded corners, fill=orange!40, shape=rectangle, minimum width=0.5cm, minimum height = 0.5cm] (v0c) at (3,0.5) {$\Lambda$};
    \node[draw, rounded corners, fill=orange!40, shape=rectangle, minimum width=0.5cm, minimum height = 0.5cm] (v1) at (4,0.5) {$V^\dagger$};
     \node[draw, rounded corners, fill=orange!40, shape=rectangle, minimum width=0.5cm, minimum height = 0.5cm] (v2b) at (5,0.5) {$V^\dagger$};
    \node[draw, rounded corners, fill=orange!40, shape=rectangle, minimum width=0.5cm, minimum height = 0.5cm] (v00b) at (2,-0.5) {$U^\dagger$};
    \node[draw, rounded corners, fill=orange!40, shape=rectangle, minimum width=0.5cm, minimum height = 0.5cm] (v00c) at (3,-0.5) {$\Lambda^\dagger$};
    \node[draw, rounded corners, fill=orange!40, shape=rectangle, minimum width=0.5cm, minimum height = 0.5cm] (v11) at (4,-0.5) {$V$};
     \node[draw, rounded corners, fill=orange!40, shape=rectangle, minimum width=0.5cm, minimum height = 0.5cm] (v22b) at (5,-0.5) {$V$};
    \draw [thick] (v0b) -- (v0c)
    (v0c) -- (v1)
    (v1) -- (v2b)
    (v0b) -- (v00b)
    (v0c) -- (v00c)
    (v1) -- (v11)
    (v2b) -- (v22b)
    (v00b) -- (v00c)
    (v00c) -- (v11)
    (v11) -- (v22b);
    \draw [thick] (3,0.75) -- (3,1);
    \draw [thick] (3,-0.775) -- (3,-1);
    \node [] (t2) at (5.85,0) {$=$};
     \node[draw, rounded corners, fill=orange!40, shape=rectangle, minimum width=0.5cm, minimum height = 0.5cm] (v0d) at (7,0.5) {$\Lambda$};
    \node[draw, rounded corners, fill=orange!40, shape=rectangle, minimum width=0.5cm, minimum height = 0.5cm] (v00d) at (7,-0.5) {$\Lambda^\dagger$};
    \draw [thick] (v0d) -- (v00d);
    \draw [thick] (v0d) to [bend left=90] (v00d);
    \draw [thick] (v0d) to [bend right=90] (v00d);
    \draw [thick] (7,0.75) -- (7,1);
    \draw [thick] (7,-0.775) -- (7,-1);
    \end{tikzpicture}} $\qquad$
    \subfigure[]{\centering
    \begin{tikzpicture}
    \node[draw, rounded corners, fill=orange!40, shape=rectangle, minimum width=0.5cm, minimum height = 0.5cm] (v0b) at (2,0.5) {$U$};
    \node[draw, rounded corners, fill=orange!40, shape=rectangle, minimum width=0.5cm, minimum height = 0.5cm] (v0c) at (3,0.5) {$\Lambda$};
    \node[draw, rounded corners, fill=orange!40, shape=rectangle, minimum width=0.5cm, minimum height = 0.5cm] (v1) at (4,0.5) {$V^\dagger$};
     \node[draw, rounded corners, fill=orange!40, shape=rectangle, minimum width=0.5cm, minimum height = 0.5cm] (v2b) at (5,0.5) {$V^\dagger$};
    \node[draw, rounded corners, fill=orange!40, shape=rectangle, minimum width=0.5cm, minimum height = 0.5cm] (v00b) at (2,-1) {$U^\dagger$};
    \node[draw, rounded corners, fill=orange!40, shape=rectangle, minimum width=0.5cm, minimum height = 0.5cm] (v00c) at (3,-1) {$\Lambda^\dagger$};
    \node[draw, rounded corners, fill=orange!40, shape=rectangle, minimum width=0.5cm, minimum height = 0.5cm] (v11) at (4,-1) {$V$};
    \node[draw, rounded corners, fill=orange!40, shape=rectangle, minimum width=0.5cm, minimum height = 0.5cm] (v22b) at (5,-1) {$V$};
    \node[draw, rounded corners, fill=blue!40, shape=rectangle, minimum width=0.5cm, minimum height = 0.5cm] (vint) at (3,-0.25) {$\sigma_Z$};
    \draw [thick] (v0b) -- (v0c)
    (v0c) -- (v1)
    (v1) -- (v2b)
    (v0b) -- (v00b)
    (v1) -- (v11)
    (v0c) -- (vint)
    (vint) -- (v00c)
    (v2b) -- (v22b)
    (v00b) -- (v00c)
    (v00c) -- (v11)
    (v11) -- (v22b);
    \draw [thick] (3,0.75) -- (3,1);
    \draw [thick] (3,-1.275) -- (3,-1.5);
    \node [] (t2) at (5.85,-0.25) {$=$};
    \node[draw, rounded corners, fill=orange!40, shape=rectangle, minimum width=0.5cm, minimum height = 0.5cm] (v0d) at (7,0.5) {$\Lambda$};
    \node[draw, rounded corners, fill=orange!40, shape=rectangle, minimum width=0.5cm, minimum height = 0.5cm] (v00d) at (7,-1) {$\Lambda^\dagger$};
    \node[draw, rounded corners, fill=blue!40, shape=rectangle, minimum width=0.5cm, minimum height = 0.5cm] (vintb) at (7,-0.25) {$\sigma_Z$};
    \draw [thick] (v0d) to [bend right=90] (v00d)
    (v0d) -- (vintb)
    (vintb) -- (v00d)
    (v0d) to [bend left=90] (v00d);
    \draw [thick] (7,0.75) -- (7,1);
    \draw [thick] (7,-1.275) -- (7,-1.5);
    \end{tikzpicture}} \\
\end{tabular}
\end{center}
\caption{Diagrammatic illustration in panel (a) of the predictor defined in Eq. \eqref{eq::pred}, where the variational tensor $W$ is expressed as a Matrix Product State (MPS) in the mixed-canonical form, as introduced in Eq. \eqref{eq::Wmps} \cite{SCHOLLWOCK201196}. The inclusion of an orthogonality center $\Lambda$ enables efficient tensor contractions over repeated indices $s_1, s_2, \dots, s_N$, for the pictorial purpose of a $N=4$ lattice: (b) the reduced density matrix in label space, and (c) the local magnetization at each site. \label{fig::mps}}
\end{figure}

A key advantage of MPS lies in their controllable approximation: by tuning the bond dimension $\chi$, one can balance the trade-off between computational cost and model expressiveness. This scalability, combined with their transparency and interpretability, makes MPS an ideal candidate for integrating quantum-inspired representations into classical machine learning pipelines.

We leverage a quantum-inspired representation to train a variational object, referred to as a quantum mask $W$, capable of performing multi-class classification on classical input data. Each input array $\mathbf{x} \in \mathbb{R}^N$ is mapped to a quantum state $\ket{\Phi(\mathbf{x})}$ in the exponentially large Hilbert space $\mathbb{R}^{2^N}$, using a local feature encoding scheme:
\begin{equation}
\mathbf{x} \in \mathbb{R}^N \mapsto \ket{\Phi(\mathbf{x})} = \bigotimes_{j=1}^{N} \ket{\phi(x^{(j)})} \in \mathbb{R}^{2^N},
\end{equation}
where each sample $x^{(j)}$ is embedded into a single-qubit state given by $\ket{\phi(x^{(j)})}=\left( \cos(x^{(j)}), \ \sin(x^{(j)}) \right)^\intercal$ \cite{stoudenmire2017supervisedlearningquantuminspiredtensor}. This encoding partially preserves the geometric structure of the input while allowing quantum-inspired models to act linearly in Hilbert space but nonlinearly in input space, thus enhancing their expressive power.

To enable efficient learning and scalability, we represent the quantum mask $W$ using a MPS ansatz, which efficiently parameterizes high-dimensional tensors by factorizing them into a sequence of low-rank matrices, with controllable expressive capacity governed by the bond dimension $\chi$. The construction relies on a truncated singular value decomposition (SVD) at each bipartition of the one dimensional system $M=U \Sigma V^\dagger$, where we consider the top $\chi$ singular values to limit the growth of entanglement and computational cost \cite{SCHOLLWOCK201196}.

The MPS decomposition of the mask $W$ takes the form \cite{SCHOLLWOCK201196}:
\begin{equation} \label{eq::Wmps}
    W^\ell_{s_1,\dots,s_N} = \sum_{a_1,\dots,a_N} U_{s_1,a_1} U_{s_2,a_2}^{a_1} \dots U_{s_{j-1},a_{j-1}}^{a_{j-2}}  \Lambda_{s_j}^{a_{j-1},\ell,a_j} V_{s_{j+1},a_j}^{\dagger a_{j+1}} \dots V_{s_{N-1},a_{N-2}}^{\dagger a_{N-1}} V_{s_N,a_{N-1}}^{\dagger}
\end{equation}
where the tensor $\Lambda$, located at site $j$, acts as the orthogonality center. This structure, shown schematically in Fig. \ref{fig::mps}(a), allows for efficient contraction and differentiation, even in high-dimensional input spaces \cite{PhysRevLett.124.037201, evenbly2022}.

For a classification task involving classes labeled by $\ell$, the MPS defines a predictive model that acts on the encoded product state:
\begin{equation} \label{eq::pred}
    f^\ell_W (\mathbf{x}) = \sum_{s_1 \dots s_N} W^\ell_{s_1,\dots,s_N} \ket{\phi(x^{(1)})^{s_1}} \otimes \dots \otimes \ket{\phi(x^{(N)})^{s_N}},
\end{equation}
where $\ket{\phi(x^{(j)})^{s_j}}$ is the single-qubit encoding of the $j$-th feature and $W^\ell$ denotes the MPS tensor network associated with class $\ell$. During training, the parameters of $W$ are optimized to minimize the mean squared error cost function:
\begin{equation} \label{eq::cost}
    \mathcal{C}(W) = \frac{1}{2} \sum_{\omega=1}^{N_T} \sum_\ell \left( f^\ell_{W}(\mathbf{x}_{\omega}) - y^\ell_{\omega} \right)^2,
\end{equation}
where $N_T$ is the number of training samples $(\mathbf{x}_{\omega}, y^\ell_{\omega})$, and $y^\ell_{\omega}=1$ if sample $\omega$ belongs to class $\ell$, and zero otherwise. The predicted label is determined by the component showing the highest absolute value $|f^\ell_{W}(\mathbf{x})|$ \cite{chen2023machinelearningtreetensor}. 

The learning process is driven by the minimization of the task-specific cost function in Eq. \eqref{eq::cost}, typically designed to penalize misclassification. To update the MPS parameters, we employ a two-site gradient descent scheme, which enables the optimization of adjacent tensors simultaneously while preserving the canonical form of the MPS
\begin{center}
\begin{tikzpicture}
    \node[draw, rounded corners, fill=orange!40, shape=rectangle, minimum width=0.5cm, minimum height = 0.5cm] (v0b) at (2,0.5) {$U$};
    \node[draw, rounded corners, fill=orange!40, shape=rectangle, minimum width=0.5cm, minimum height = 0.5cm] (v0c) at (3,0.5) {$\Lambda$};
    \node[draw, rounded corners, fill=orange!40, shape=rectangle, minimum width=0.5cm, minimum height = 0.5cm] (v1) at (4,0.5) {$V^\dagger$};
     \node[draw, rounded corners, fill=orange!40, shape=rectangle, minimum width=0.5cm, minimum height = 0.5cm] (v2b) at (5,0.5) {$V^\dagger$};
    \draw [thick] (v0b) -- (v0c)
    (v0c) -- (v1)
    (v1) -- (v2b);
    \draw [thick] (3,0.75) -- (3,1.25);
    \draw [thick] (5,0.225) -- (5,-0.25);
    \draw [thick] (2,0.25) -- (2,-0.25);
    \draw [thick] (3,0.25) -- (3,-0.25);
    \draw [thick] (4,0.225) -- (4,-0.25);
    \node [] (t3) at (2.65,1.25) {$\ell$};
    \node [] (t2) at (2.7,-0.15) {$s_j$};
    \node [] (t5) at (4.5,-0.15) {$s_{j+1}$};
    \node [] (t2) at (5.8,0.5) {$=$};
    \node[draw, rounded corners, fill=orange!40, shape=rectangle, minimum width=0.5cm, minimum height = 0.5cm] (v0bB) at (6.5,0.5) {$U$};
    \node[draw, rounded corners, fill=orange!40, shape=rectangle, minimum width=1.5cm, minimum height = 0.5cm] (v0cB) at (8,0.5) {$B$};
     \node[draw, rounded corners, fill=orange!40, shape=rectangle, minimum width=0.5cm, minimum height = 0.5cm] (v2bB) at (9.5,0.5) {$V^\dagger$};
    \draw [thick] (v0bB) -- (v0cB)
    (v0cB) -- (v2bB);
    \draw [thick] (8,0.75) -- (8,1.25);
    \draw [thick] (9.5,0.225) -- (9.5,-0.25);
    \draw [thick] (6.5,0.25) -- (6.5,-0.25);
    \draw [thick] (7.5,0.25) -- (7.5,-0.25);
    \draw [thick] (8.5,0.25) -- (8.5,-0.25);
    \node [] (t3B) at (7.65,1.25) {$\ell$};
    \node [] (t2B) at (7.2,-0.15) {$s_j$};
    \node [] (t5B) at (9,-0.15) {$s_{j+1}$};
\end{tikzpicture}
\end{center}
referring to a single right step, which leads to a sweep when back at the initial sites pair. The update rule requires the introduction of the following quantity \\
\begin{center}
    \begin{tikzpicture}
    \node[draw, rounded corners, fill=orange!40, shape=rectangle, minimum width=0.5cm, minimum height = 0.5cm] (v0b) at (1,0.5) {$U$};
    \node[draw, rounded corners, fill=orange!40, shape=rectangle, minimum width=0.5cm, minimum height = 0.5cm] (v2b) at (4,0.5) {$V^\dagger$};
    \node[draw, fill=purple!40, shape=circle, inner sep=5 pt] (v00) at (1,-0.75) {};
    \node[draw, fill=purple!40, shape=circle, inner sep=5 pt] (v00b) at (3,-0.75) {};
    \node[draw, fill=purple!40, shape=circle, inner sep=5 pt] (v22b) at (2,-0.75) {};
    \node[draw, fill=purple!40, shape=circle, inner sep=5 pt] (v22c) at (4,-0.75) {};
    \draw [thick] (v0b) -- (v00)
    (v2b) -- (v22c);
    \draw [thick] (2,0) -- (2,-0.5);
    \draw [thick] (3,0) -- (3,-0.5);
    \draw [thick] (1.27,0.5) -- (1.5,0.5);
    \draw [thick] (3.4,0.5) -- (3.65,0.5);
    \node [] (t2) at (1.7,-0.15) {$s_j$};
    \node [] (t5) at (3.5,-0.15) {$s_{j+1}$};
    \node [] (t2) at (4.8,0) {$=$};
    \node[draw, fill=purple!80, shape=circle, inner sep=5 pt] (v00) at (5.5,-0.75) {};
    \node[draw, fill=purple!40, shape=circle, inner sep=5 pt] (v00b) at (7.5,-0.75) {};
    \node[draw, fill=purple!40, shape=circle, inner sep=5 pt] (v22b) at (6.5,-0.75) {};
    \node[draw, fill=purple!80, shape=circle, inner sep=5 pt] (v22c) at (8.5,-0.75) {};
    \draw [thick] (6.5,0) -- (6.5,-0.5);
    \draw [thick] (7.5,0) -- (7.5,-0.5);
    \draw [thick] (5.5,0) -- (5.5,-0.5);
    \draw [thick] (8.5,0) -- (8.5,-0.5);
    \node [] (t2) at (6.2,-0.15) {$s_j$};
    \node [] (t5) at (8,-0.15) {$s_{j+1}$};
    \node [] (t3) at (10,0) {\( \displaystyle = \quad \ket{\widetilde{\Phi}(\mathbf{x})} \)};
\end{tikzpicture}
\end{center}
equivalently expressed in the predictor
\begin{equation}
    f^\ell_W (\mathbf{x}) = \sum_{s_j, s_{j+1}} \sum_{a_{j-1},a_{j+1}} B^{a_{j-1},\ell,a_{j+1}}_{s_j,s_{j+1}} \ket{\widetilde{\Phi}(\mathbf{x})^{s_j, s_{j+1}}_{a_{j-1},a_{j+1}}},
\end{equation}
such that
\begin{equation}
    \Delta B^\ell = -\frac{\partial \mathcal{C}}{\partial B^\ell} = \sum_{\omega = 1}^{N_T} \ket{\widetilde{\Phi}(\mathbf{x}_{\omega})} \otimes (y^\ell_{\omega} - f^\ell_{W}(\mathbf{x}_{\omega})).
\end{equation}

At each step of the training process, the variational tensor $B^\ell$ is updated via a gradient descent rule governed by the learning rate $\alpha$, as follows:
\begin{equation}
    B'^\ell=B^\ell + \alpha \Delta B^\ell.
\end{equation}
Given that the evolution induced by gradient descent is inherently non-unitary, we must renormalize the updated quantum state to maintain physical interpretability and numerical stability. This is accomplished by compressing $B'^\ell$ to the prescribed bond dimension $\chi$ through a truncated singular value decomposition (SVD), followed by a global rescaling. Effectively, this post-selection protocol implements a projective operation onto the dominant subspace spanned by the retained singular vectors, analogous to a measurement collapsing the state onto its most significant modes \cite{wiersema2023}.

To monitor the learning dynamics and interpret the internal structure of the quantum mask $W$, we evaluate a set of physically meaningful observables:
\begin{enumerate}
	\item reduced density matrix $\varrho^{(\ell)}$ in the label space, extracted using the contraction scheme shown in Fig. \ref{fig::mps}(b). This quantity encodes the coherence and distinguishability between label states during training;
	\item local magnetization $\braket{\sigma_Z^{k,i}}$, computed separately for each label $k=0,1,2$ and feature index $i = 1,\dots,N$, as depicted in Fig. \ref{fig::mps}(c). These expectation values reveal the contribution of individual input features to the classification decision, serving as a form of interpretable attribution.
\end{enumerate}

Following each tensor update, we perform a two-site SVD:
\begin{equation}
    B'_{s_i, s_{i+1}}=U_{s_i} \Sigma V^\dagger_{s_{i+1}},
\end{equation}
where $\Sigma$ contains the singular values $\{\lambda_j\}_{j=1,\dots,\chi}$. Depending on the direction of the optimization sweep, the updated orthogonality center $\Lambda$ is obtained by contracting $\Sigma$ with $U$ (leftward) or $V^\dagger$ (rightward). The singular value spectrum encodes entanglement information across the bond between sites $i$ and $i+1$, allowing us to compute the von Neumann entanglement entropy:
\begin{equation}
    S(i)=-\sum_{j=1}^\chi \lambda_j^2 \log \lambda_j^2
\end{equation}
This quantity measures the degree of quantum correlation (or representational complexity) between the subsystems to the left and right of the bond, serving as an informative diagnostic of model capacity and learning phase transitions.

Furthermore, to analyze how class-specific information is distributed across the network, we perform independent SVDs on the updated tensors $B'^{0}_{s_i, s_{i+1}}$ and $B'^{1}_{s_i, s_{i+1}}$, corresponding to the label-resolved components of the quantum mask. This enables the computation of label-specific entanglement entropies $S^{0}(i)$ and $S^{1}(i)$, providing a nuanced view of how each class utilizes correlations across the lattice, a key indicator of information localization.

\subsection{Transfer Entropy}

Causal relations among quantum masks during training sweeps are monitored by means of transfer entropy \cite{schreiber_te}
\begin{equation}
    TE_{\ell,j \rightarrow \ell',i} = \sum p\left(\braket{\sigma^{\ell',i}_Z(t+1)},\braket{\sigma^{\ell',i,(\tau')}_Z(t)},\braket{\sigma^{\ell,j,(\tau)}_Z(t)}\right) \frac{p\left(\braket{\sigma^{\ell',i}_Z(t+1)}\big\vert \braket{\sigma^{\ell',i,(\tau')}_Z(t)},\braket{\sigma^{\ell,j,(\tau)}_Z(t)}\right)}{p\left(\braket{\sigma^{\ell',i}_Z(t+1)}\big\vert \braket{\sigma^{\ell',i,(\tau')}_Z(t)}\right)},
\end{equation}
where $\braket{\sigma^{\ell,j,(\tau)}_Z(t)}=(\braket{\sigma^{\ell,j}_Z(t)},\dots,\braket{\sigma^{\ell,j}_Z(t-\tau+1)})$ with $t$ and $\tau$ integer labels that identify the training sweep and delay, respectively. For our numerical purposes, it is useful to express the transfer entropy in terms of conditional entropies between processes $\Sigma^{\ell,j}_Z$ yielding the sequences of expectation values $\braket{\sigma^{\ell,j}_Z(t)}$, particularly in the entropy difference form exploited by the Kraskov-St\"ogbauer-Grassberger nearest-neighbor method \cite{ksg_method,frenzel,varley} described in Appendix \ref{appA}
\begin{align}
    TE_{\ell,j \rightarrow \ell',i} &= H\left(\Sigma^{\ell',i}_Z(t+1)\big\vert \Sigma^{\ell',i,(\tau')}_Z(t)\right) - H\left(\Sigma^{\ell',i}_Z(t+1)| \Sigma^{\ell',i,(\tau')}_Z(t),\Sigma^{\ell,j,(\tau)}_Z(t)\right), \nonumber \\
    &= I\left(\Sigma^{\ell',i}_Z(t+1), \Sigma^{\ell,j,(\tau)}_Z(t)\big\vert \Sigma^{\ell',i,(\tau')}_Z(t)\right)
\end{align}
that is, the reduction in uncertainty of the future of $\Sigma^{\ell',i}_Z$ due to knowledge of past $\Sigma^{\ell,j}_Z$, beyond what is already known from past $\Sigma^{\ell',i}_Z$. In the following we will considered averaged transfer entropy over spin pairs $TE_{\ell \rightarrow \ell'} = \frac{1}{N^2} \sum_{i,j} TE_{\ell,j \rightarrow \ell',i}$.

For a given spin pair $i, j$ and masks pair $\ell, \ell'$, we have a time series of expectation values with length equal to the number of training sweeps. Corresponding to each training step we build $(\braket{\sigma^{\ell',i}_Z(t+1)}, \braket{\sigma^{\ell',i,(\tau')}_Z(t)}, \braket{\sigma^{\ell,j,(\tau)}_Z(t)})$ and we determine the distance $\varepsilon_k(t)$ to the $k$-th nearest neighbor with respect to the maximum norm. Then we switch to the marginals $(\braket{\sigma^{\ell',i}_Z(t+1)}, \braket{\sigma^{\ell',i,(\tau')}_Z(t)})$, $(\braket{\sigma^{\ell',i}_Z(t+1)}, \braket{\sigma^{\ell,j,(\tau)}_Z(t)})$, and $\braket{\sigma^{\ell',i}_Z(t+1)}$ in order to determine for each case the number of points with distance strictly less than $\varepsilon_k(t)$, denoted by $N_{\ell' \ell'}$, $N_{\ell' \ell}$ and $N_{\ell'}$ \cite{frenzel, faes}
\begin{equation}
    TE_{\ell,j \rightarrow \ell',i} = \psi(k) + \braket{\psi(N_{\ell'}+1) - \psi(N_{\ell' \ell'}+1) - \psi(N_{\ell' \ell}+1)},
\end{equation}
where $\psi$ is the digamma function and $\braket{\cdot}$ denotes the average over all time steps. We choose to adopt the number of nearest neighbors $k=4$.

\subsection{O-information}

The O-information is a high-order information theory quantity designed to quantify the dominant mode of interaction among multiple variables in a system. It was introduced as ``information about organizational structure'', and it coincides with interaction information for three variables \cite{rosas}. As an extension of mutual information beyond pairwise relationships, it distinguishes whether a group of variables primarily shares redundant or synergistic information. A positive O-information indicates that variables tend to be redundant, meaning their dependencies are largely shared or overlapping—often pointing to coordinated, predictable behavior. Conversely, a negative O-information signals synergy, where information is distributed across variables in a way that cannot be fully captured by subsets, thus implying more complex, complementary interactions. This makes O-information particularly well-suited for analyzing emergent dependencies in learning systems, such as in quantum or classical neural networks, where one seeks to track the transition from disordered, synergistic regimes to structured, redundant ones.

We consider the three processes associated with predicted labels in output, namely $f_W^{\ell}(\mathbf{x})$ with $\ell=0,1,2$. At each training step these scores are sampled from the aforementioned processes $F^{\ell}_W(t)$, such that we express the O-information, or equivalently, the interaction information
\begin{align} \label{eq:oinfo}
    \Omega\left(F^0_W(t), F^1_W(t), F^2_W(t) \right) & = I\left(F^0_W(t), F^1_W(t), F^2_W(t)\right) \\
    & = I\left(F^0_W(t), F^1_W(t) \right) - I\left(F^0_W(t), F^1_W(t) \big\vert F^2_W(t) \right). \nonumber 
\end{align}
where the result is invariant under any swap in the variables order. These terms are separately evaluated \cite{ksg_method}
\begin{align}
    I\left(F^0_W(t), F^1_W(t) \right) & = \psi(k) + \psi(T) - \braket{\psi(N_{F^0}+1) + \psi(N_{F^1}+1)}, \\
    I\left(F^0_W(t), F^1_W(t) \big\vert F^2_W(t) \right) & = \psi(k) + \braket{\psi(N_{F^2}+1) - \psi(N_{F^0 F^2}+1) - \psi(N_{F^1 F^2}+1)},
\end{align}
where $N_{F^0}, N_{F^1}, N_{F^2}, N_{F^0 F^2}, N_{F^1 F^2}$ are the number of points within the $k$-th nearest neighbor distance for each marginal and $T$ is the total number of training steps.

\section{Results} \label{sec3}

\subsection{Classification performances}

\begin{figure}[t!]
\includegraphics[width=\linewidth]{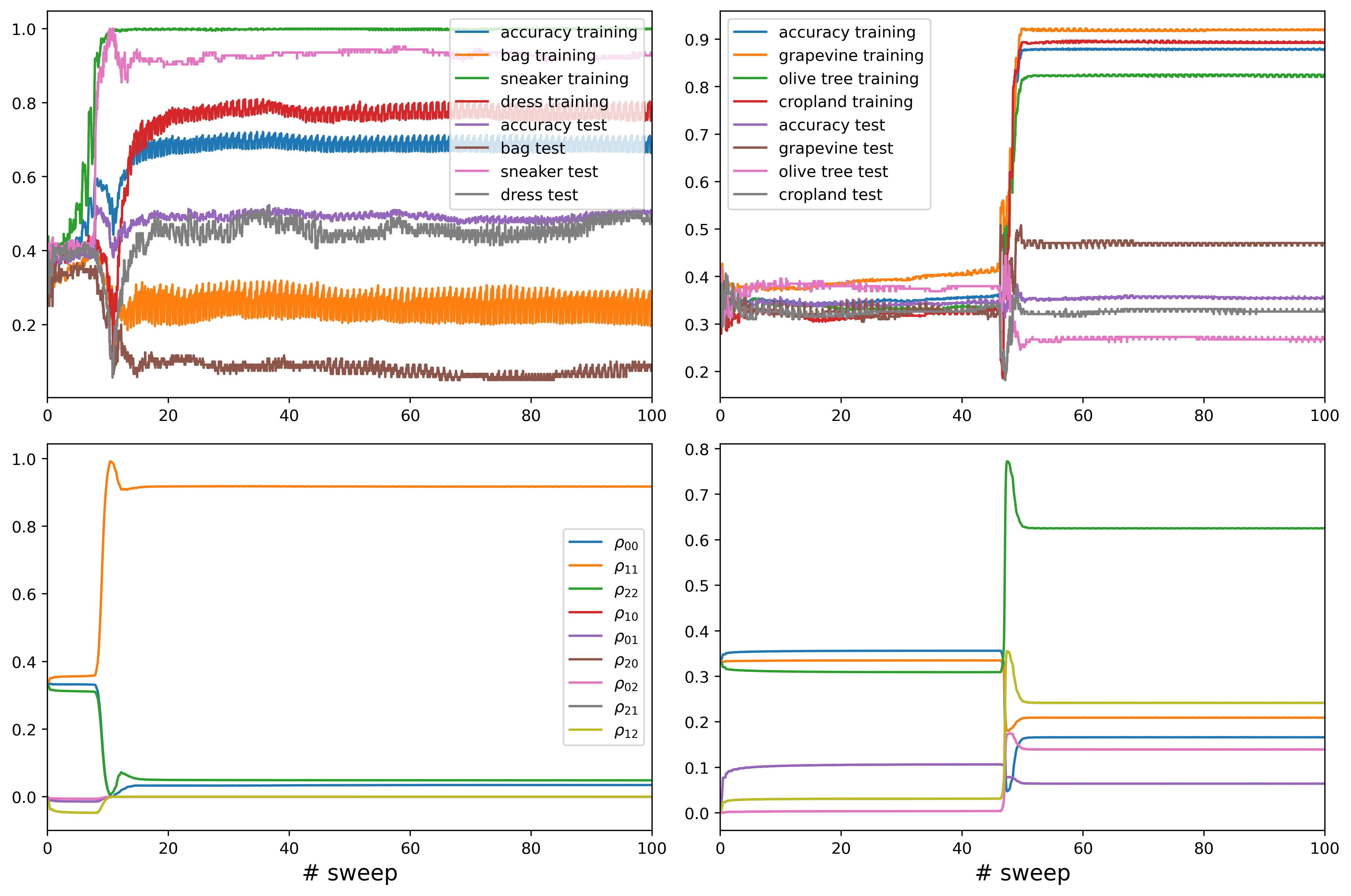}
\caption{Performance analysis of two quantum MPS classifiers on 3-class problems: Fashion MNIST (left) and hyperspectral land cover classification (right). Top panels show overall and per-class training/test accuracies. Bottom panels display the evolution of the reduced density matrix elements in the label subspace. \label{fig:metrics}}
\end{figure}

The performance of the MPS classifier on the two multiclass problems is resumed in the top panels of Figure \ref{fig:metrics}. Each task involves three classes as introduced in Section \ref{sec2}:
\begin{enumerate}
    \item Fashion MNIST: dress, sneaker, bag;
    \item Hyper-spectral land cover: cropland, olive tree, grapevine.
\end{enumerate}

The top panels display evaluation metrics, while the bottom panels show the evolution of the reduced density matrix in the label subspace for each case. This provides insight into how the quantum classifier organizes information in the output space.

In the top left panel the performance on fashion MNIST is characterized in terms of the overall training and test accuracy, plotted alongside the per-class performance. Sneaker class achieves the highest performance, with the test accuracy stabilizing around $90\%$. Dress class follows with a slightly lower performance.These patterns yield a grokking transition based on an effective features extraction useful for unseen data. Bag class performs poorly as a confounding case, with training accuracy around $25\%$, with oscillations, and test performance reaching a value even smaller than $10\%$.

The general training accuracy increases steadily and saturates near $70\%$, but the test accuracy is limited by the poor generalization of the bag class.
Oscillations and performance disparities suggest insufficient data quality for certain classes imposed by image rescaling. The MPS classifier handles sneakers and dresses well, likely due to distinguishable features consisting in horizontally and vertically elongated shapes in rescaled pixel space. However, bags are harder to distinguish.

In the top right panel, the performance on hyper-spectral land cover classification is shown. There is a sharp transition in accuracy around sweep 45, indicating a shift in the overfitting regime, since the test performances are low. Grapevine achieves the highest training accuracy (around $90\%$) and test accuracy around $50\%$. Olive tree performs moderately, with almost $80\%$ training and $35\%$ test accuracy. Cropland performs the worst, with lowest test accuracy around $28\%$. The generalization gap is more pronounced than in the fashion MNIST task.

In bottom panels the reduced density matrix in the label subspace is characterized. On the left for fashion MNIST, the evolution of the $3 \times 3$ reduced density matrix shows diagonal elements ($\varrho_{00}, \varrho_{11}, \varrho_{22}$) representing class populations:
$\varrho_{11}$ corresponding to sneaker saturates around 0.95, while $\varrho_{00}$ and $\varrho_{22}$ are much lower. Off-diagonal elements (e.g. $\varrho_{12}, \varrho_{01}$) decrease and stabilize near zero, so the model reaches a nearly classically separable form in the label subspace. Off-diagonal suppression indicates vanishing label entanglement, meaning that the model makes confident, distinct predictions for each input. On the right for land cover classification, a sudden jump near sweep 45 aligns with the accuracy jump. After this transition the diagonal elements stabilize to less balanced values (e.g. $\varrho_{00}$ around $0.6$, others $0.2$), while off-diagonal elements (e.g. $\varrho_{12}, \varrho_{01}$) remain significant, some up to $0.2$. Unlike the fashion MNIST task, the land cover model retains quantum coherence in the label subspace. This suggests that the classifier operates in a superposed decision space, potentially due to the overlap in spectral features between the vegetation types.

\subsection{Magnetization pattern extraction}

\begin{figure}[t!]
\centering
\includegraphics[width=\linewidth]{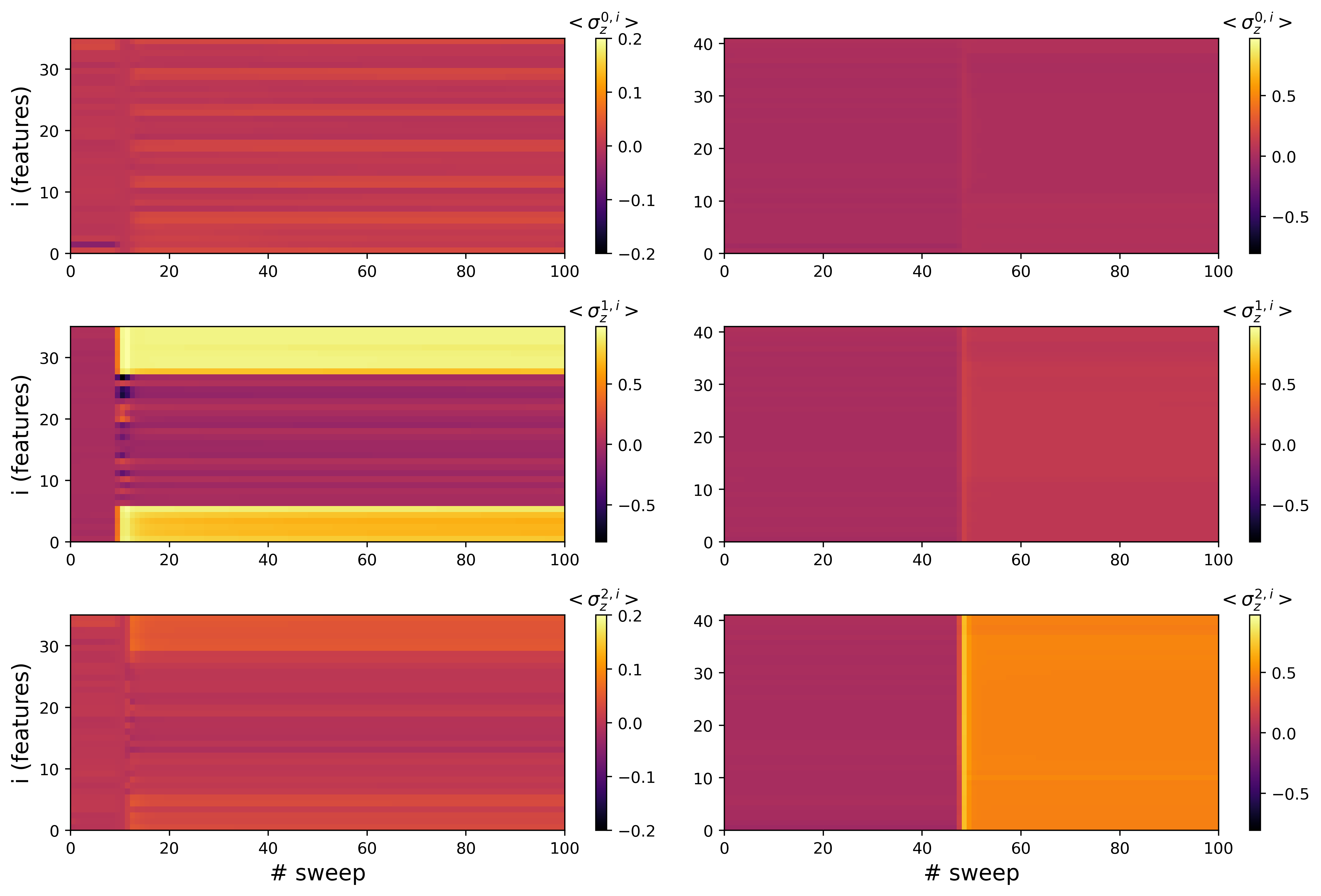}
\caption{Local Z-magnetization $\braket{\sigma^{\ell,i}_Z}$ of the trained quantum masks for the three class labels (rows) in fashion MNIST (left) and hyper-spectral land cover classification (right). Each row corresponds to the following label: first row refers to dress / cropland, second row to sneaker / olive tree, and third row to bag / grapevine. \label{fig:magnetization}}
\end{figure}

The evolution of the local magnetization $\braket{\sigma^{\ell,i}_Z}$ along the $Z$ direction is represented in Figure \ref{fig:magnetization} for each site $i$ (feature index) and for each of the three label-targeted quantum masks ($\ell = 0, 1, 2$) across training sweeps (horizontal axis). The left column refers to the fashion MNIST classification task, while the right column corresponds to the hyper-spectral land cover classification.

Each row represents one of the three class-specific quantum masks:
\begin{enumerate}
    \item top row: label 0 - dress / cropland;
    \item middle row: label 1 - sneaker / olive tree;
    \item bottom row: label 2 - bag / grapevine.
\end{enumerate}

The three quantum masks for the fashion MNIST dataset show distinct, label-specific magnetization patterns across features. Each mask develops structure early during training (around sweep 10), where specific feature regions acquire non-trivial magnetization values. Label 1 (middle panel) shows the earliest and most pronounced activation, with magnetizations saturating near $+ 1$ on subsets of features. This aligns with the high classification performance of the sneaker class. Label 0 (top panel) follows shortly after, with a visible delay of a few sweeps relative to label 1. Its final magnetization values are smaller in magnitude and less structured. Label 2 (bottom panel) develops more slowly and remains less polarized, consistent with the weaker classification performance for bag.

The masks for fashion MNIST evolve asynchronously and specialize at different times. The slight temporal offset between the activation of label 1 (sneaker) and label 0 (dress) reflects the model’s ease in learning separable features for sneakers earlier than for dresses. Bag features remain harder to encode, with less magnetization contrast.

The right column refers to the  land cover classification, where all three quantum masks undergo a sharp and synchronized transition in their local magnetizations at approximately sweep 50. This transition is abrupt and coherent across the three labels, resulting in simultaneous polarization across many features. Before sweep 50, magnetizations are close to zero for all classes, indicating the model is in an unstructured or pre-learning phase. After sweep 50, all masks acquire relatively uniform and structured magnetization patterns, though with subtle differences in magnitude.

\begin{figure}[t!]
\centering
\includegraphics[width=\linewidth]{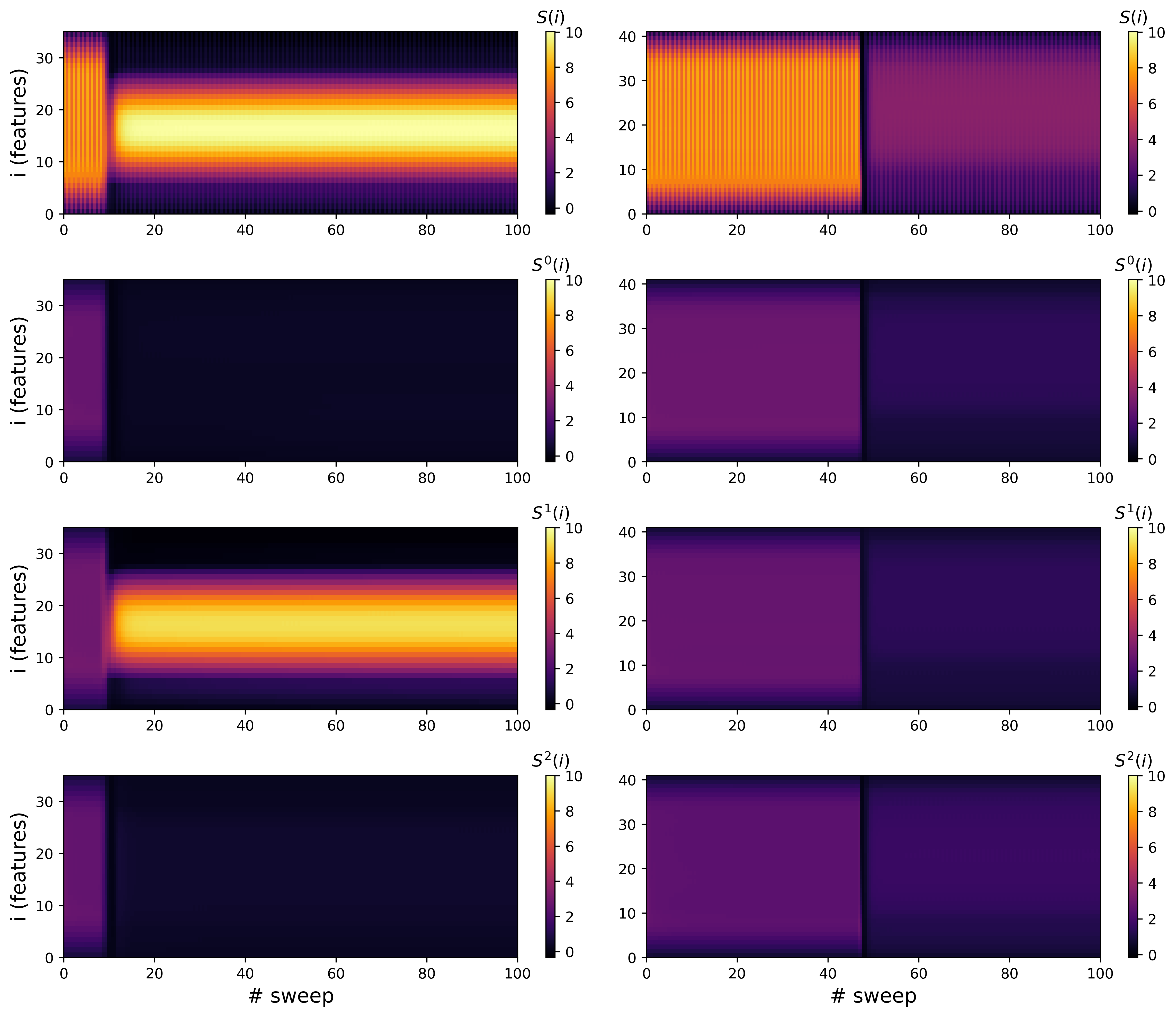}
\caption{Entanglement entropy $S(i)$ along the MPS chain during training for fashion MNIST (left) and land cover classification (right). The top row shows total entropy, while rows 2-4 display label-resolved components: label 0 for dress/cropland, label 1 for sneaker/olive tree, label 2 for bag/grapevine. \label{fig:entropy}}
\end{figure}

Unlike the fashion MNIST task, the masks for land cover classification display collective behavior, with a global learning transition likely triggered by a structural modification in the training process. The absence of delay between label activations suggests that the model required a global capacity boost to start encoding meaningful class distinctions, reflecting the presence of an overfitting regime, as shown in Figure \ref{fig:metrics}.

Figure \ref{fig:entropy} illustrates the evolution of the entanglement entropy along the MPS chain during training sweeps. The vertical axis corresponds to the site index $i$ (i.e. feature index) and each panel shows the local bipartite von Neumann entropy, capturing the degree of entanglement across a cut in the MPS at the edge between site $i$ and $i+1$.

Left column refers to the fashion MNIST task, while the right column to the hyper-spectral land cover classification. The top row describes total entanglement entropy $S(i)$ for the full output state. From the second up to the fourth row we have label-specific contributions $S^{\ell}(i)$, ordered as follows: dress / cropland, sneaker / olive tree, bag / grapevine.

In both tasks, the top row clearly displays a transition from a high-entropy, volume law regime at early sweeps to a lower-entropy, sub-volume law regime as training progresses \cite{pomarico2025grokkingentanglementtransitiontensor}. This is typical of MPS initialization with random parameters, where entanglement is initially extensive across the chain. As training optimizes the tensor network, the system moves toward a more structured state that requires less entanglement, indicating emergent efficient interactions among features required to minimize the cost function.

Initially, for fashion MNIST all positions exhibit high entanglement, consistent with a volume law state. As training progresses (especially by sweep around 10), entanglement becomes highly localized. The third row, corresponding to label 1 (sneaker), dominates the entropy dynamics. It maintains high entropy centered along the chain, indicating non-trivial internal correlations within the label-1 subspace. After the transition, two regions with high local magnetization become weakly entangled, consistent with effective feature segmentation. The other labels (0 = dress, 2 = bag) show negligible entropy after early sweeps, indicating that those subspaces become effectively decoupled from the evolving state. The MPS optimization concentrates most of the entanglement in the subspace associated with label 1 (sneaker), suggesting that this class is the dominant attractor in the learned representation.

The initial state for land cover classification exhibits a broad, uniform volume law entanglement across the MPS chain as well. At sweep around 50, a sharp drop in entropy is observed, consistent with a collective transition already seen in the magnetization and accuracy plots. After the transition, all labels exhibit low entanglement across the chain, with no particular label dominating the entropy budget. The near-flat, low-entropy pattern reflects the model’s attempt to suppress complexity and extract spatially coarse features. Unlike fashion MNIST, the learning here leads to an overall reduction in entanglement across all class subspaces, consistent with overfitted data representations.

\subsection{Causal information transfer between quantum masks}

\begin{figure}[t!]
\includegraphics[width=\linewidth]{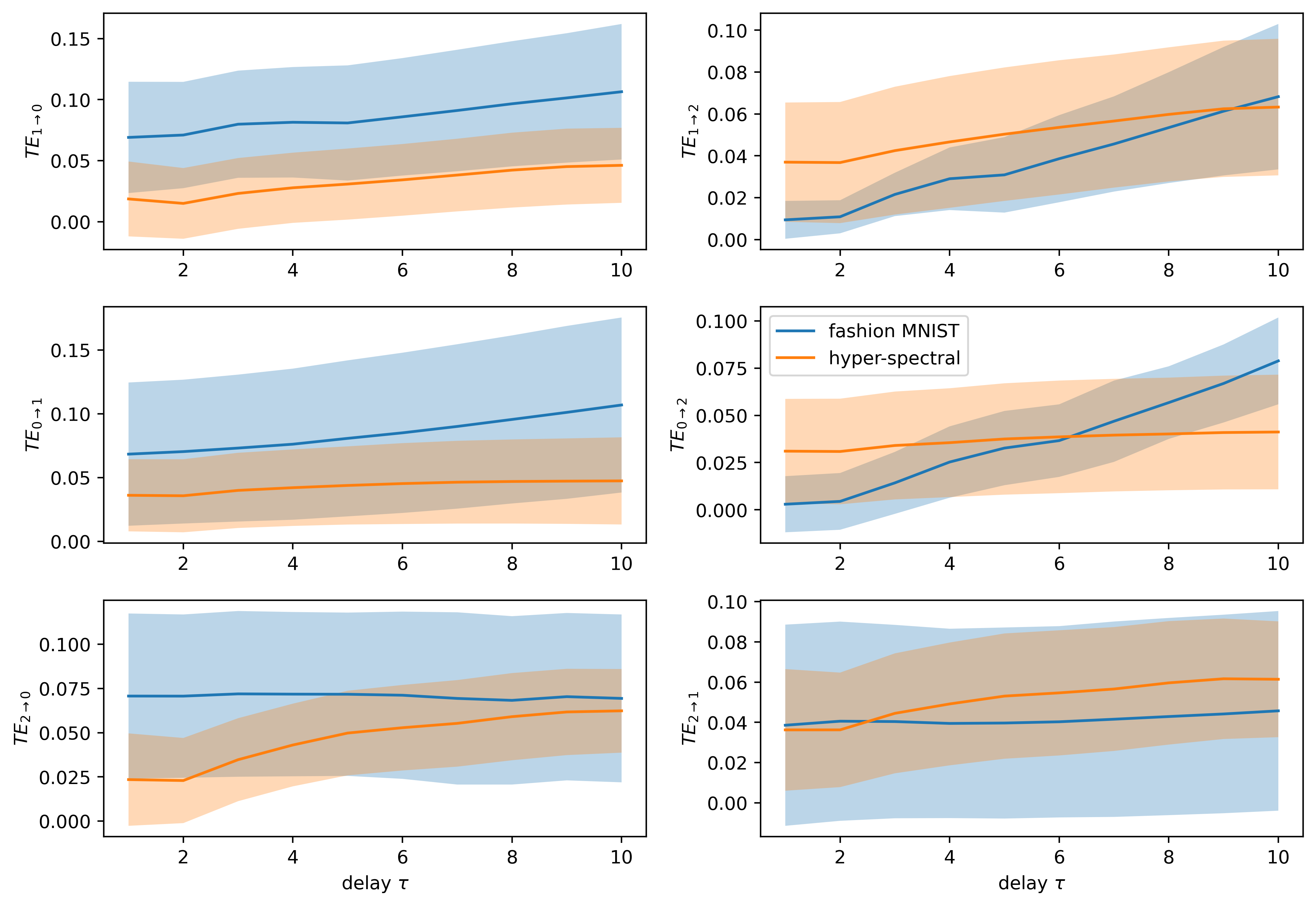}
\caption{Transfer entropy between pairs of quantum masks as a function of training delay $\tau$, comparing fashion MNIST (blue) and hyperspectral (orange) classification tasks. Each panel shows the directional TE from one label-specific mask to another, averaged over spin pairs. Solid lines are referred to the mean value, while shaded regions determine one standard deviation. \label{fig:te}}
\end{figure}

The proposed detection of grokking in the presence of an entanglement transition is based on the determination of causal relations between quantum masks during the training dynamics. Classical transfer entropy is a measure of information flow originally framed for time series. Figure \ref{fig:te} presents this quantity between each quantum mask pair as a function of time delay $\tau$ (in training sweeps, ranging from 1 to 10) for the two previously discussed classification tasks: fashion MNIST (blue) and hyperspectral land cover (orange). Each subplot represents the directional transfer entropy from one label-specific quantum mask to another, averaged over all pairs of spins $Z$ expectation values between the source and target masks. Shaded regions determine one standard deviation around the mean.

In the top-left panel the information transfer from label 1 (sneaker / olive) to label 0 (dress / cropland) is endowed with a clear and persistent separation between the fashion MNIST and hyper-spectral cases. The blue and orange curves are separated by more than one standard deviation, indicating statistically significant causal influence from mask 1 to mask 0 in fashion MNIST. The separation peaks around $\tau=2, 3$ suggest that the influence of the sneaker mask on the dress mask in the MPS is strongest at short delays during training. This is corroborated by the z-scores for $TE_{1 \rightarrow 0}$ in the Table \ref{tab1}, i.e. $\tau = 2$ with $z=1.515$ and $\tau = 3$ with $z=1.524$. The first value further increases to 2.783 when averaging is restricted to strongly magnetized spins, i.e., spins encoding the most relevant features.

The remaining directions for information transfer could be signaled by $TE_{1 \rightarrow 2}$, $TE_{0 \rightarrow 2}$ showing moderate differences between tasks, but the statistical separation significantly weakens with increasing delay. This is reflected in the decreasing z-scores, e.g., $TE_{1 \rightarrow 2}$ drops from 1.3 to nearly 0 by $\tau=10$. $TE_{2 \rightarrow 1}$, $TE_{2 \rightarrow 0}$ and $TE_{0 \rightarrow 1}$ show overlapping shaded regions, with low z-scores and weak directional dependencies. These directions appear less informative in distinguishing dynamics between the two tasks.
For the hyperspectral case, transfer entropy values remain consistently lower and flatter, suggesting that information flow between masks is weaker and less temporally structured.

\begin{table}[b!]
\caption{Z-scores collected from comparisons of quantum masks transfer entropy mean and standard deviation in Figure \ref{fig:te}.\label{tab1}}
		\begin{tabular}{ccccccc}
			\toprule
			$\tau$	& $TE_{1 \rightarrow 0}$	& $TE_{1 \rightarrow 2}$     & $TE_{0 \rightarrow 1}$  & $TE_{0 \rightarrow 2}$ & $TE_{2 \rightarrow 0}$  & $TE_{2 \rightarrow 1}$ \\
1 & 1.298 & 1.304 & 0.726 & 1.265 & 1.248 & 0.056  \\
2 & 1.515  & 1.221 & 0.774 & 1.177 & 1.297 & 0.107 \\
3 & 1.524  & 0.918 & 0.725 & 0.850 & 1.005 & 0.102 \\
4 & 1.415  & 0.713 & 0.726 & 0.421 & 0.782 & 0.244 \\
5 & 1.279   & 0.750 & 0.763 & 0.192 & 0.597 & 0.333 \\
6 & 1.296 & 0.554 & 0.800 & 0.079 & 0.489 & 0.358  \\
7 & 1.290  & 0.396 & 0.857 & 0.283 & 0.366 & 0.369 \\
8 & 1.287  &  0.215 & 0.934 & 0.662 & 0.243 & 0.409  \\
9 & 1.295   &  0.036 & 1.009 & 1.005 & 0.230 & 0.429  \\
10 & 1.344  &  0.148 & 1.098 & 1.399 & 0.187 & 0.387  \\
		\end{tabular}
\end{table}

In fashion MNIST, the strong $TE_{1 \rightarrow 0}$ signal implies that the learning dynamics of the sneaker mask (label 1) drives or conditions those of the dress mask (label 0). This supports earlier observations that label 1 dominates both magnetization and entropy dynamics.
The asymmetry of transfer entropy confirms a causal hierarchy: the sneaker mask organizes meaningful features early, and this structure propagates to influence other masks. In hyper-spectral classification, the flat transfer entropy curves point to a more synchronized or globally entangled evolution, with no single mask strongly conditioning the others.

This analysis demonstrates that transfer entropy between quantum masks can reveal asymmetric causal dependencies in the learning process. For fashion MNIST, label 1 (sneaker) emerges as a dynamical driver, with $TE_{1 \rightarrow 0}$ being the most statistically robust and interpretable signature. In contrast, the hyperspectral task exhibits a more balanced and collective mask interaction, lacking clear causal dominance. These findings provide a dynamical fingerprint of how quantum models internally organize information during training, potentially guiding task-specific architecture adjustments or mask pruning strategies.

\subsection{Scores redundancy peak at grokking}

\begin{figure}[t!]
\includegraphics[width=\linewidth]{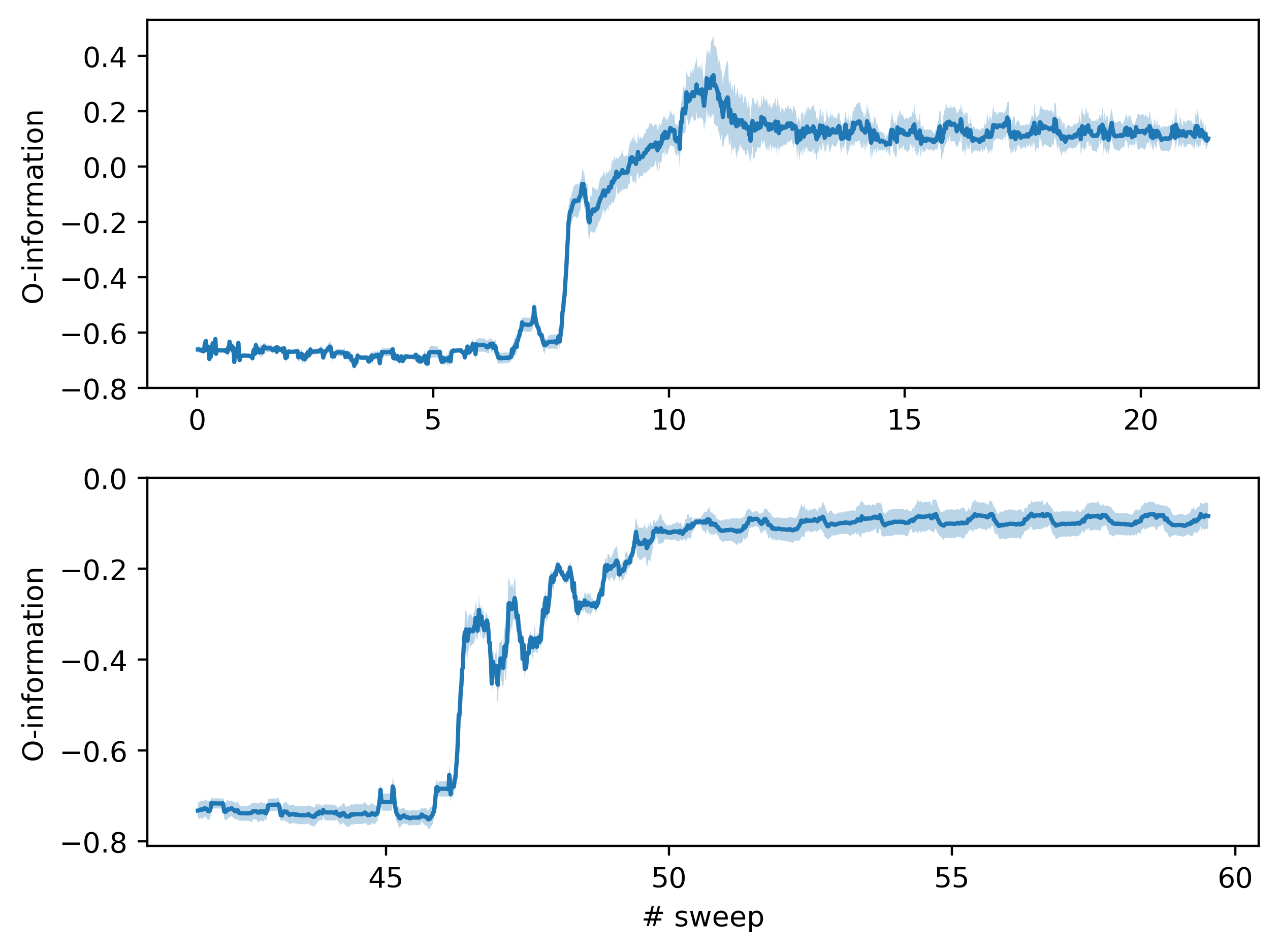}
\caption{O-information among the output scores of the three label-specific quantum masks during training for fashion MNIST (top) and hyperspectral (bottom) classification tasks. Both panels focus on sweeps hosting the entanglement transition, potentially related with grokking. Solid lines and shaded regions are referred to mean and standard deviation of the six permutations available for variables order in Eq. \eqref{eq:oinfo}. \label{fig:oinfo}}
\end{figure}

A proper characterization of quantum masks dynamics in targeted multiclass classification tasks resides in evaluating O-information among probabilistic output scores. Figure \ref{fig:oinfo} shows the evolution of the O-information averaged during training, over the available order permutations of the variables in the Eq. \eqref{eq:oinfo}, for the two different tasks: fashion MNIST (top panel) and hyperspectral land cover classification (bottom panel). The O-information is a high-order information-theoretic quantity that captures whether correlations among variables are redundant (positive values) or synergistic (negative values).

The initial regime is driven by the random MPS in both tasks: the training begins with strongly negative O-information, indicating a regime where synergy dominates the interactions among the label scores. This is expected from the randomly initialized MPS, where the output probabilities for each class are not yet structured and exhibit complex, non-redundant joint behavior. This is characteristic of a random classifier where different class predictions are not yet disentangled.

Corresponding to the entanglement transition, leading to grokking in the fashion MNIST task (top panel), a sharp transition occurs around sweep 8-9. The O-information jumps into the positive regime, peaking significantly before stabilizing. This transition coincides with earlier observations of improved accuracy, magnetization structure, and entanglement collapse. The shift to positive O-information reflects a regime dominated by redundancy. In contrast, the hyperspectral task (bottom panel) shows a more gradual transition around sweep 46, without reaching positive values. The O-information remains negative even after the transition, suggesting that the learned representations maintain synergistic, entangled relationships more typical of a random or overfitted classifier. This aligns with previous observations of weak causal mask interactions, low label-specific magnetization polarization, and flat transfer entropy curves. In both cases it is possible to identify a moderate peak forecasting the transition jump during the previous MPS sweep.

The O-information offers a compact global measure of the nature of inter-dependencies in the model output: a positive O-information (redundancy) is provided by predictive outputs highly coordinated and mutually informative, a signature of successful learning, while a negative O-information (synergy) among predictive outputs expresses complementary, distributed information. The clear divergence between the two tasks highlights the qualitative difference in learning behavior: fashion MNIST exhibits a grokking transition, with outputs reorganizing toward redundancy and predictability, the hyperspectral model, even after fitting, retains features of a disordered system, with scores synergy, indicating limited generalization.

This analysis of O-information complements earlier metrics (accuracy, entropy, magnetization, transfer entropy), revealing that fashion MNIST undergoes a coherent learning transition to redundancy-dominated dynamics, while hyperspectral classification remains trapped in a synergistic, less interpretable regime. The O-information thus acts as a powerful high-order probe for emergent structure and model interpretability in QML.

\section{Discussion} \label{sec4}

Our study investigates the training dynamics of MPS-based quantum machine learning models in the context of three-class classification problems, extending previous analyses that focused on binary tasks such as those presented in \cite{pomarico2025grokkingentanglementtransitiontensor}. By comparing fashion MNIST and hyperspectral satellite imagery datasets, we reveal qualitatively distinct behaviors in how these models generalize, organize internal correlations, and evolve information structure throughout training.

A key finding is the identification of a grokking-like phase transition in the three-classes fashion MNIST task, marked by a sharp improvement in test accuracy, an abrupt drop in entanglement entropy from the initial volume law regime, and a peak in O-information, signaling the onset of redundancy-driven learning. This aligns with theoretical models of grokking in classical deep learning, where generalization follows a delay after fitting, driven by the reorganization of internal representations \cite{power2022grokkinggeneralizationoverfittingsmall, liu2022understandinggrokkingeffectivetheory, liu2023grokkingcompressionnonlinearcomplexity, miller2024grokkingneuralnetworksempirical,  huang2024unifiedviewgrokkingdouble,varma2023explaininggrokkingcircuitefficiency,rubin2024grokkingorderphasetransition, clauw2024informationtheoreticprogressmeasuresreveal}. In contrast, the hyperspectral task, despite overfitting training data, fails to exhibit this transition. It maintains persistent synergistic correlations and weak directional information flow between class-specific quantum masks, resembling the behavior of a random classifier.

The selection of the PRISMA hyper-spectral dataset and the fashion MNIST image dataset reflects a deliberate choice to investigate learning dynamics across two qualitatively distinct scenarios. Fashion MNIST serves as a structured benchmark in which class-defining features, such as shape, are spatially coherent and separable. However, among the three selected classes, i.e. dress, sneaker, and bag, the last functions as a confounding class, sharing visual features with both of the others and thereby introducing ambiguity into the classification task. This setup allows us to explore how the model resolves feature competition and reorganizes its internal representations across training. In contrast, the PRISMA dataset represents a significantly more complex and noisy regime. Its hyper-spectral data are high-dimensional, with subtle spectral differences between land-cover types, i.e. grapevine, olive tree and cropland, and potential ambiguities due to mixed pixels and acquisition noise. These factors hinder generalization and trap the model in overfitting learning phases. By juxtaposing these two datasets, we can contrast structured learning with non-learning behavior and interpret emergent information-theory signatures, such as redundancy, synergy, and causal flow, across fundamentally different data regimes.

The methodological choice in our study was the focus on three-class classification problems, which serves as the minimal setting for exploring both high-order statistical dependencies and causal interactions within the learning dynamics. Unlike binary classification, which limits the scope of interaction analysis to pairwise relations \cite{pomarico2025grokkingentanglementtransitiontensor}, the three-class scenario allows us to investigate how multiple output channels co-evolve and influence each other during training. In particular, this structure enables the application of O-information distinguishing between redundancy, where variables share overlapping information, and synergy, where the whole carries more information than the sum of its parts. By applying this formalism to the classification scores associated with the three output classes, we gain insight into how the network restructures its internal representations across the grokking transition. Moreover, having three label-specific MPS masks makes it possible to examine directional dependencies in the form of transfer entropy, offering a window into the causal relationships that emerge as different subspaces magnetize. 

The application of transfer entropy to monitor causal interactions between masks and the use of O-information to distinguish synergy from redundancy offer new diagnostic tools for probing the internal structure of quantum-inspired learners. This richer structure, unavailable in simpler binary setups, proves essential for revealing the interplay between information organization, learning dynamics, and entanglement restructuring in quantum-inspired models. These methods go beyond accuracy metrics, providing interpretable signatures of training phases and class-specific dynamics, and may be particularly valuable for understanding overfitting and generalization in high-dimensional regimes.

Our results underscore the importance of task complexity and input structure. Despite pixel rescaling, the fashion MNIST data allows the MPS model to develop a clear causal hierarchy among masks (e.g., the sneaker class driving the dress one), while the spectral noise of hyperspectral imagery appears to suppress such organization. This suggests that data structure, not model architecture alone, governs the emergence of interpretable learning dynamics.

Looking forward, these findings motivate several directions. First, refining the role of redundancy and synergy in multiclass classification may guide architecture design, such as adaptive bond dimensions or class-specific entanglement constraints. Second, the integration of classical and quantum representations (e.g., through hybrid or shallow quantum circuits) could leverage the observed causal flows to reduce training overhead or compress information flow. Finally, the application of this framework to other structured datasets, such as time-series, geospatial patterns, or biological signals, may further test the generality of grokking and provide deeper insight into the informational geometry of learning in quantum systems.

\section{Conclusions}

In this work, we explored the training dynamics of quantum-inspired MPS classifiers in three-class learning tasks, focusing on fashion MNIST and hyperspectral land cover classification. By combining conventional performance metrics with high-order information-theory tools, namely transfer entropy and O-information, we uncovered distinct learning behaviors between the two tasks. Fashion MNIST exhibited a clear grokking transition characterized by redundancy, entanglement compression, and directional information flow among class-specific quantum masks. In contrast, the hyperspectral model remained in a synergistic regime, indicative of overfitting and weak internal structure. These results demonstrate the utility of causal and high-order information measures for diagnosing learning phases and generalization in quantum machine learning, paving the way for more interpretable and task-adaptive quantum models.

\section{Acknowledgements}

D.P. acknowledges the support by PNRR MUR project CN00000041-``National Center for Gene Therapy and Drugs based on RNA Technology''. S.S. was supported by the project ``Higher-order complex systems modeling for personalized medicine,'' Italian Ministry of University and Research (funded by MUR, PRIN 2022-PNRR, code P2022JAYMH, CUP: H53D23009130001). Authors were supported by the Italian funding within the ``Budget MIUR - Dipartimenti di Eccellenza 2023 - 2027'' (Law 232, 11 December 2016) - Quantum Sensing and Modelling for One-Health (QuaSiModO), CUP:H97G23000100001. Authors want to thank the project ``Genoma mEdiciNa pERsonalizzatA –GENERA'', local project code T3-AN-04 – CUP H93C22000500001, financed under the Health Development and Cohesion Plan 2014-2020, Trajectory 3 ``Regenerative, predictive and personalized medicine'' - Action line 3.1 ``Creation of a precision medicine program for the mapping of the human genome on a national scale'', referred to in the Notice of the Ministry of Health published in the Official Journal no. 46 of 24 February 2021. Authors want to thank the Funder: Project funded under the National Recovery and Resilience Pan (NRRP), Mission 4 Component 2 Investment 1.4 - Call for tender No. 3138 of 16 December 2021 of Italian Ministry of University and Research funded by the European Union - NextGenerationEU (award number/project code: CN00000013), and Concession Decree No. 1031 of 17 February 2022 adopted by the Italian Ministry of University and Research (CUP: D93C22000430001), Project title: ``National Centre for HPC, Big Data and Quantum Computing''. This work was also funded by the Italian Ministry of Enterprises and Made in ITaly (MIMIT) with the ‘``Project CALLIOPE - Casa dell’Innovazione per il One Health’'' (FSC 2014–2020, CUP: E53C22002800001) and with the WADIT project (Water Digital Twin) co-funded by the Italian Ministry of Enterprises and Made in ITaly (MIMIT), ``Decreto n. Accordi per l'innovazione di cui al D.M. 31 dicembre 2021 e D.D. 18 marzo 2022'' project n. F/310272/01-05/X56. Computational resources were provided by ReCaS Bari \cite{recas}. The fashion MNIST and PRISMA datasets that support the findings of this study are openly available in Kaggle at \url{https://www.kaggle.com/datasets/zalando-research/fashionmnist} and through authentication \url{https://www.asi.it/en/earth-science/prisma/}, respectively.

\appendix
\renewcommand\thefigure{\thesection.\arabic{figure}}

\section[\appendixname~\thesection]{Kraskov-St\"ogbauer-Grassberger method \label{appA}}

The Kraskov-St\"ogbauer-Grassberger nearest-neighbor method is based on the definition of Shannon entropy of a continuous random variable $X$, in a metric space endowed with distance $||x-x'||$, as an average of the ``surprise'' $\log \frac{1}{\mu(x)}$ with respect to the related probability measure \cite{varley}, such that we can build the estimator
\begin{equation} \label{eq:entropy}
    H(X) = -\frac{1}{T} \sum_{i=1}^T \widehat{\log \mu} (x_i),
\end{equation}
with random sample $(x_1, \dots, x_T)$ of $T$ realizations of $X$. For each point $x_i$ we aim to determine the probability $P_k(\varepsilon) \mathrm{d}\varepsilon$ that there is one point within distance $r \in \left[\frac{\varepsilon}{2},\frac{\varepsilon+\mathrm{d}\varepsilon}{2} \right]$, such that there are $k-1$ points at smaller distances and $T-k-1$ at larger distances. Since we will estimate such probabilities locally, the condition $k \ll T$ has to be satisfied. The targeted density $\mu(x)$ imposes a mass $p_i(\varepsilon) = \int_{|| x - x_i ||<\varepsilon/2} \mathrm{d}x \ \mu(x)$ centered at $x_i$, such that the probability of getting successful detection of the $k$-th nearest neighbor exploits the trinomial coefficient $\binom{T-1}{1,k-1,T-k-1}$ as follows
\begin{equation}
    P_k(\varepsilon) \mathrm{d}\varepsilon = \frac{(T-1)!}{(k-1)! (T-k-1)!} (1-p_i)^{T-k-1} p_i^{k-1}\mathrm{d}p_i,
\end{equation}
with normalization $\int_0^\infty P_k(\varepsilon) \mathrm{d}\varepsilon = 1$ ensured by iteratively integrating by parts to obtain $\int_0^1 (1-p_i)^{T-k-1} p_i^{k-1}\mathrm{d}p_i = \frac{(k-1)!(T-k-1)!}{(T-1)!}$. The expectation value of $\log p_i$ is evaluated over the remaining $T-1$ points with respect to $x_i$ according to \cite{ksg_method}
\begin{align}
    \mathbb{E} \left[ \log p_i \right] & = \int_0^\infty \mathrm{d} \varepsilon \ P_k(\varepsilon) \ \log p_i(\varepsilon) = \frac{(T-1)!}{(k-1)! (T-k-1)!} \int_0^1 \mathrm{d} p \ (1-p)^{T-k-1} p^{k-1} \log p \nonumber \\
    & = \psi(k) - \psi(T),
\end{align}
where $\psi(z)$ is the digamma function. 

Under the assumption of constant $\mu(x)$ in the neighborhood of $x_i$ and according to the choice of maximum norm presented in Section \ref{sec2}, it is possible to deduce the estimator $\widehat{\log \mu} (x)$ by exploiting $p_i(\varepsilon) \approx \varepsilon^d \mu(x_i)$, where $d$ is the dimension of $x$. In this way we obtain the final expression for Eq. \eqref{eq:entropy}
\begin{equation} \label{eq:estimate}
    H(X) = -\psi(k) + \psi(T) + \frac{d}{T} \sum_{i=1}^T \log \varepsilon(i).
\end{equation}

The estimation of the mutual information takes into account the joint random variable $W=(X,Y)$. We consider for each $w_i$, $i=1,\dots,T$, the distance $\varepsilon/2$ to the $k$-th nearest neighbor, endowed with a probability $P_k(\varepsilon)$. The dimension of $w$ is $2d$, such that
\begin{equation}
    H(X,Y) = -\psi(k) + \psi(T) + \frac{2d}{T} \sum_{i=1}^T \log \varepsilon(i).
\end{equation}
In order to estimate marginal entropies, we have that Eq. \eqref{eq:estimate} holds true for any $k$ and it has not to be fixed. If there are $N_x(i)$ points within the distance $\varepsilon(i)$ around $x_i$, chosen such that the $N_x(i)+1$-th neighbor lies at $x=x_i \pm \varepsilon(i)/2$, we can express
\begin{equation}
    H(X)=-\braket{\psi(N_x+1)} + \psi(T) + \frac{d}{T} \sum_{i=1}^T \log \varepsilon(i),
\end{equation}
as well as for $H(Y)$ with $N_y$ respectively. The mutual information then reads \cite{ksg_method}
\begin{equation}
    I(X,Y) = \psi(k) + \psi(T) - \braket{\psi(N_x+1) + \psi(N_y+1)}.
\end{equation}
The last quantity exploited in our analysis is the conditional mutual information
\begin{equation}
    I(X,Y|Z) = H(X,Z) + H(Y,Z)-H(X,Y,Z)-H(Z),
\end{equation}
where by keeping into account that the dimension of $(x,y,z)$ is $3d$, we obtain \cite{frenzel}
\begin{equation}
    I(X,Y|Z) = \psi(k) + \braket{\psi(N_z+1) - \psi(N_{xz}+1) - \psi(N_{yz}+1)}.
\end{equation}


\end{document}